\documentclass[useAMS,usenatbib]{mn2e}

\usepackage{graphicx}

\newcommand\rs[1]{_\mathrm{#1}}
\def\gsim{\;\lower4pt\hbox{${\buildrel\displaystyle >\over\sim}$}\,}

\title[Modelling the 2014 outburst of the nova V745 Sco]
{Origin of asymmetries in X-ray emission lines
from the blast wave of the 2014 outburst of nova V745~Sco}

\author[S. Orlando et al.]{Salvatore Orlando$^{1}$\thanks{E-mail: orlando@astropa.inaf.it},
 Jeremy J. Drake$^{2}$, Marco Miceli$^{1,3}$\\ \\
$^{1}$INAF - Osservatorio Astronomico di Palermo ``G.S. Vaiana'',
       Piazza del Parlamento 1, 90134, Palermo, Italy\\
$^{2}$Harvard-Smithsonian Center for Astrophysics, 60 Garden
              Street, Cambridge, MA 02138, USA\\
$^{3}$Dip. di Fisica \& Chimica, Universit\`a di Palermo,
       Piazza del Parlamento 1, 90134, Palermo, Italy\\
}
\begin{document}

\date{Accepted. Received}

\pagerange{\pageref{firstpage}--\pageref{lastpage}} \pubyear{}

\maketitle

\label{firstpage}

\begin{abstract}
The symbiotic nova V745~Sco was observed in outburst on 2014 February 
6. Its observations by the {\it Chandra} X-ray Observatory at days
16 and 17 have revealed a spectrum characterized by asymmetric and
blue-shifted emission lines. Here we investigate the origin of these
asymmetries through three-dimensional hydrodynamic simulations
describing the outburst during the first 20 days of evolution. The 
model takes into account thermal conduction and radiative cooling
and assumes a blast wave propagates through an equatorial density 
enhancement. From the simulations, we synthesize the X-ray emission
and derive the spectra as they would be observed with {\it Chandra}.
We find that both the blast wave and the ejecta distribution are
efficiently collimated in polar directions due to the presence of
the equatorial density enhancement. The majority of the X-ray
emission originates from the interaction of the blast with the
equatorial density enhancement and is concentrated on the equatorial
plane as a ring-like structure. Our ``best-fit'' model requires a
mass of ejecta in the outburst $M\rs{ej} \approx 3\times
10^{-7}\,M_{\odot}$ and an explosion energy $E\rs{b} \approx 3
\times 10^{43}$~erg and reproduces the distribution of emission
measure vs temperature and the evolution of shock velocity and
temperature inferred from the observations.  The model predicts
asymmetric and blue-shifted line profiles similar to those observed
and explains their origin as due to substantial X-ray absorption
of red-shifted emission by ejecta material. The comparison of
predicted and observed Ne and O spectral line ratios reveals no
signs of strong Ne enhancement and suggests the progenitor is a CO
white dwarf.
\end{abstract}


\section{Introduction}
\label{s:intro}

V745~Sco is a symbiotic nova that was observed in its latest outburst
on 2014 February 6 (\citealt{2014AAN...497....1W}). Previous outbursts
were recorded in 1937 and 1989 (\citealt{1989Msngr..58...34D}) and,
probaly, one was missed in the 1960s (\citealt{2010yCat..21870275S}).
This makes of V745~Sco a member of the elusive group of recurrent
novae (\citealt{1989Msngr..58...34D}). The characteristics of the
V745~Sco stellar system are poorly known due to the crowded galactic
bulge region in which it sits; it is thought to be a close binary,
comprising a red giant star and a white dwarf
(\citealt{1989Msngr..58...34D}) with an orbital period of $510\pm
10$~days, and located at a distance of $7.8\pm 1.2$~kpc
(\citealt{2010yCat..21870275S}). In this class of
objects, material is transferred from the companion red giant onto
the surface of the white dwarf. The outbursts occur on the white
dwarf when the transferred material reaches sufficient temperature
and density to trigger a thermonuclear runaway (e.g.
\citealt{1999A&A...344..177A, 2000NewAR..44...81S}).

The 2014 outburst was monitored since the early phases of its
evolution by an intensive observing campaign, including observations
ranging from radio, to X-ray and $\gamma$-ray wavelengths. A summary
of the observations is presented in \cite{2015MNRAS.454.3108P}.
Observations of the {\it Chandra} X-ray Observatory Transmission
Grating Spectrometers on UT 2014 February 22, and 23 (namely between
15.8 and 17.4 days since the outburst) revealed a rich spectrum of
emission lines indicative of emitting plasma with temperatures
ranging between a few MK and tens of MK (\citealt{2016ApJ...825...95D}).
The spectral analysis has shown that X-ray line profiles are
significantly asymmetric and too strongly peaked to be explained
by a spherically-symmetric blast wave (\citealt{2016ApJ...825...95D});
also the lines present a systematic blueshift of $160\pm 10$~km~s$^{-1}$.
All these features have been interpreted as evidence of significant
blast collimation in analogy with the findings of previous studies
of other nova outbursts.

In recent years, there has been a growing consensus in the
literature that blast collimation is a common feature of nova
outbursts. For instance, collimation signatures analogous to those
found for V745~Sco have been found during the 2006 outburst of the
recurrent nova RS~Oph at radio, infrared, optical, and X-ray
wavelengths (e.g. \citealt{2006Natur.442..276S, 2006Natur.442..279O,
2006ApJ...652..629B, 2009ApJ...707.1168L, 2009ApJ...691..418D}).
Convincing theoretical support of the idea that nova blasts are
highly collimated has been provided by accurate multi-dimensional
hydrodynamic models. These have shown that the interaction of the
explosion with either an accretion disc or a disc-like equatorial
density enhancement (hereafter EDE) around the white dwarf produces a
characteristic bipolar shock morphology in which both the blast and
the ejecta from the outburst are strongly collimated in polar
directions (e.g. \citealt{2008A&A...484L...9W, 2009A&A...493.1049O,
2010ApJ...720L.195D, 2012MNRAS.419.2329O, 2015ApJ...806...27P}).
In the case of the 2006 outburst of RS~Oph, \cite{2009A&A...493.1049O}
have shown that the broadening of emission lines observed with {\it
Chandra} is the result of the interaction of the blast wave with
an EDE and their asymmetries are due to substantial X-ray absorption
of red-shifted emission by ejecta material. Besides that,
ascertaining the presence of an EDE in these systems is important
also to unveil the origin of $\gamma$-ray emission which seems to
originate at the interface between the equatorial and polar regions
(\citealt{2014Natur.514..339C, 2015MNRAS.450.2739M}) where, most
likely, the blast interacts with the EDE.

In this paper, we explore the effects of the EDE and the red
giant companion on the blast wave morphology and ejecta distribution
during the 2014 outburst of V745~Sco through hydrodynamic modeling.
The aims include: 1) to ascertain the role of the EDE in the
collimation of blast wave and ejecta during this particular event;
2) to provide a deeper insight on the origin of the asymmetries,
broadening, and blueshifts revealed in the profiles of X-ray emission
lines; and 3) to constrain the environment surrounding this binary
system by deriving the average density structure and geometry of the
circumstellar medium (CSM) immediately surrounding the nova. The
latter point may provide important clues on the final stages of
stellar evolution. Also it is relevent for our understanding of the
origin of non-thermal (synchrotron) emission observed in nova
outbursts (e.g. \citealt{2014Natur.514..339C}) which is likely due
to interaction of blast and ejecta with the dense material of the
EDE. In our approach, we synthesize the X-ray emission from the
hydrodynamic simulations and derive the spectra as they would be
observed with the {\it Chandra} Transmission Grating Spectrometers;
finally we compare the model results with observations.  In
Section~\ref{s:obsanal} we describe the hydrodynamic model, the
numerical setup, and the synthesis of X-ray emission; in Section
\ref{s:discuss} we discuss the results; and finally in Section
\ref{s:conclusion} we draw our conclusions.

\section{Hydrodynamic Modeling}
\label{s:obsanal}

The three-dimensional hydrodynamic model adopted here is similar
to that of \cite{2009A&A...493.1049O} and describes the expansion
of the blast wave from the 2014 outburst of the nova V745~Sco through
the extended outer atmosphere of the companion red giant. The blast
wave is modeled by numerically solving the time-dependent fluid
equations of mass, momentum and energy conservation, including the
radiative losses from an optically thin plasma and the thermal
conduction.  The evolution of ejecta is traced by considering a
passive tracer, $C\rs{ej}$, associated with them (see
\citealt{2009A&A...493.1049O} for more details).  The calculations
are performed using FLASH, an adaptive mesh refinement multiphysics
code for astrophysical plasmas (\citealt{Fryxell2000ApJS}). The
hydrodynamic equations for compressible gas dynamics are solved
using the FLASH implementation of the piecewise-parabolic method
(\citealt{1984JCoPh..54..174C}).

\begin{figure}
  \centering
  \includegraphics[width=8truecm]{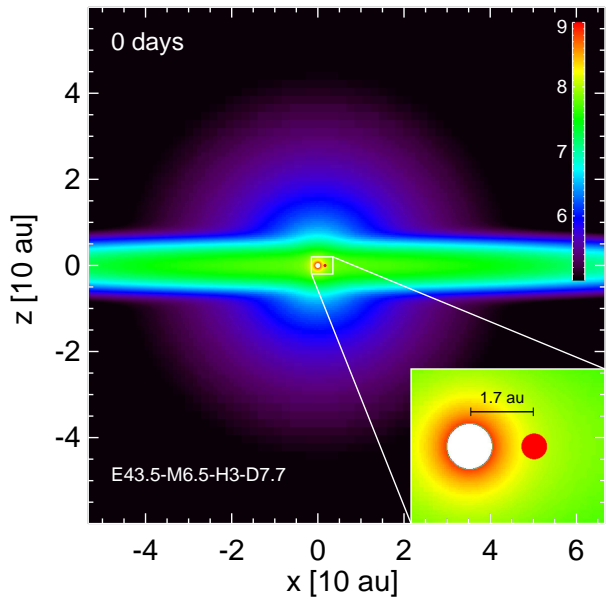}
  \caption{Colour-coded cross-section image of the gas density
  distribution, on a logarithmic scale, in units of cm$^{-3}$,
  showing the initial conditions of run E43.5-M6.5-H3-D7.7. Note
  that only one quadrant of the whole spatial domain was modeled
  numerically, namey for positive values of $y$ and $z$ (see text
  and Table~\ref{tab1}). The inset panel shows a close-up view of
  the initial geometry of the V745~Sco system. The companion star
  is at the origin (white circle on the left of the inset), and the
  initial spherical blast wave lies on the $x$ axis at $x=1.7$ au
  (red circle on the right).} \label{fig1}
\end{figure}

For the system parameters (namely binary separation, $a\rs{bs}$,
and radius of the red giant companion, $R_{*}$), we adopt the values
of \cite{2009ApJ...697..721S} (see Table~\ref{tab1}). The companion
is included as an impenetrable body with radius $R_{*} = 126\,R_{\odot}$.
We assume the gas density in the wind is proportional to $r^{-2}$
(where $r$ is the radial distance from the companion red giant) and
explore different values of the mass-loss rate. In addition to the
$r^{-2}$ density distribution, we include a diffuse, disc-like
distribution around the binary system which describes an EDE (see
Fig.~\ref{fig1}). Detailed hydrodynamic modeling studies predict
that this equatorial structure originates from gravitational
accumulation of the cool red giant wind towards the white dwarf
(e.g. \citealt{2008A&A...484L...9W, 2015ApJ...806...27P}).  Following
\cite{2012MNRAS.419.2329O}, we describe the mass density distribution
of the unperturbed CSM in Cartesian geometry as

\begin{equation}
\rho = \rho\rs{w}r\rs{pc}^{-2}+
\rho\rs{ede}~e^{[-(x/h\rs{x})^2-(y/h\rs{y})^2-(z/h\rs{z})^2]}
\label{csm}
\end{equation}

\noindent
where $\rho\rs{w} = \mu m\rs{H} n\rs{w}$ is the wind mass density
at 1~pc, $\mu \approx 1.3$ is the mean atomic mass (assuming metal
abundances of 0.5 times the solar values; \citealt{2015MNRAS.448L..35O}),
$m\rs{H}$ is the mass of the hydrogen atom, $r\rs{pc}$ is the radial
distance from the red giant in pc, $\rho\rs{ede} = \mu m\rs{H}
n\rs{ede}$ is the density of the EDE close to the red giant, and
$h\rs{x}$, $h\rs{y}$, and $h\rs{z}$ are characteristic length scales
determining the size and shape of the EDE.

Note that, in our description of the CSM, we do not consider
the effect of the white dwarf which may induce a wake like a spiral
arm due to its orbit through the red giant wind. The V745~Sco system
is believed to be quite similar to the RS~Oph system (see the
discussion at the beginning of section \ref{hydro_evol}).  Thus,
we can estimate the region expected to be dominated by spiral arms
from the work of \cite{2008A&A...484L...9W} in which the authors
reconstruct the density structure of the pre-nova CSM around RS~Oph.
From their Fig. 2, we note that the region which is heavily dominated
by the arms is within a distance of $\approx 1$~au from the white
dwarf. We expect, therefore, that our model (which neglects the
more complex structure of the CSM immediately surrounding the white
dwarf) does not describe accurately the very early (first day)
evolution of the blast. On the other hand, in this work, we aim at
comparing our model results with observations at later times ($> 1$
day), namely when the blast is expected to propagate through the
less perturbed part of the EDE. In particular we do not expect any
relevant effect on the X-ray emission synthesized at day 17,
corresponding to the time of {\it Chandra} observations.

As an initial condition, we assume a spherical Sedov-type blast
wave originating from the thermonuclear explosion with total energy
$E\rs{b}$ and ejecta mass $M\rs{ej}$. No wind phase after the
initial thermonuclear runaway event is considered (e.g.
\citealt{1994ApJ...437..802K, 2009ApJ...699.1293K}). The blast
wave is centered on the white dwarf, with an initial radius $r\rs{b0}
= 0.3$~au, offset from the red giant by 1.7~au
(\citealt{2009ApJ...697..721S}). The influence of the different
system parameters is investigated by exploring models with an initial
energy of explosion $E\rs{b}$ in the range $10^{43}-10^{44}$~erg,
ejecta mass $M\rs{ej}$ in the range $10^{-7}-10^{-6}\,M_{\odot}$,
wind density at 1 pc $n\rs{w}$ in the range $0.005-0.1$~cm$^{-3}$,
EDE density close to the red giant $n\rs{ede}$ in the range
$10^7-5\times 10^8$~cm$^{-3}$, and characteristic length scale\footnote{We
keep the values of the other length scales fixed, namely $h\rs{x}
= h\rs{y} = 54$~au.} $h\rs{z}$ in the range $1-7$~au (see
Table~\ref{tab1}). The model neglects the wind velocity ($\approx
10$~km~s$^{-1}$; e.g. \citealt{2014ApJ...785L..11B}) which is
expected to be several orders of magnitude lower than the velocity
of the blast (much larger than $1000$~km~s$^{-1}$). We consider
also additional simulations without an EDE and assuming the origin
of the blast wave to be either coincident with the origin of the
wind or offset from it by 1.7~au.  The former models are analogous
to the 1D models used by \cite{2016ApJ...825...95D} to interpret
the {\it Chandra} data and are considered here just to compare
our results with those of \cite{2016ApJ...825...95D}, whereas the
latter models highlight the effect of shielding of the blast by the
red giant secondary. For these additional simulations, we adopt
blast and wind parameters consistent with those discussed by
\cite{2016ApJ...825...95D}. The explosion and subsequent expansion
of the blast wave is followed for a total of 20~days in order to
explore the evolution of the X-ray emission till the epoch of {\it
Chandra} observations and study the effects of the circumstellar
environment on the evolution of the blast.

\begin{table*}
\caption{Adopted parameters and initial conditions for the hydrodynamic
models of the 2014 V745~Sco explosion
\label{tab1}}
\begin{center}
\begin{tabular}{llllllll}
\hline
\multicolumn{1}{l}{Parameter} & \multicolumn{4}{c}{Value} & \multicolumn{3}{l}{Notes} \\\hline
\multicolumn{1}{l}{Secondary star radius} & \multicolumn{4}{c}{$R\rs{*}=126\,R_{\odot}$} & \multicolumn{3}{l}{\cite{2009ApJ...697..721S}} \\
\multicolumn{1}{l}{Binary separation}     & \multicolumn{4}{c}{$a\rs{bs}= 1.7$ au} & \multicolumn{3}{l}{\cite{2009ApJ...697..721S}} \\
\multicolumn{1}{l}{Distance}              & \multicolumn{4}{c}{$D = 7.8$ kpc} & \multicolumn{3}{l}{\cite{2010yCat..21870275S}} \\
\multicolumn{1}{l}{Spatial domain}        & \multicolumn{4}{c}{$-50 \leq x \leq 70$ au} \\
     &   \multicolumn{4}{c}{$\hspace{0.53cm}0 \leq y \leq 60$ au} \\
     &   \multicolumn{4}{c}{$\hspace{0.53cm}0 \leq z \leq 60$ au} \\
\multicolumn{1}{l}{AMR max. resolution}   & \multicolumn{4}{c}{$2.25\times 10^{11}$ cm ($0.015$ au)} \\
\multicolumn{1}{l}{Time covered} & \multicolumn{4}{c}{0--20 days} \\ \hline
             & $E\rs{b}$ & $M\rs{ej}$ & $n\rs{w}$ & $n\rs{ede}$ & $h\rs{x}$ & $h\rs{y}$ & $h\rs{z}$\\
Model abbreviation & [erg] & $[M_\odot]$ & [cm$^{-3}$] &   [cm$^{-3}$]  & [au] & [au] & [au] \\
\hline
E43-M7-W0.1-nostar   &  $10^{43}$         & $10^{-7}$         & 0.1  &  $-$  &  $-$ &  $-$ & $-$ \\
E43.5-M7-W0.1-nostar &  $3\times 10^{43}$ & $10^{-7}$         & 0.1  &  $-$  &  $-$ &  $-$ & $-$ \\
E43-M7-W0.1          &  $10^{43}$         & $10^{-7}$         & 0.1  &  $-$  &  $-$ &  $-$ & $-$ \\
E43.3-M7-W0.1        &  $2\times 10^{43}$ & $10^{-7}$         & 0.1  &  $-$  &  $-$ &  $-$ & $-$ \\
E43-M7-H1-D8         &  $10^{43}$         & $10^{-7}$         & 0.005 &  $10^8$         & 54 & 54 & 1  \\
E43-M7-H2-D8         &  $10^{43}$         & $10^{-7}$         & 0.005 &  $10^8$         & 54 & 54 & 2  \\
E43-M7-H2-D8.7       &  $10^{43}$         & $10^{-7}$         & 0.005 &  $5\times 10^8$ & 54 & 54 & 2  \\
E43.3-M7-H2-D8       &  $2\times 10^{43}$ & $10^{-7}$         & 0.005 &  $10^8$         & 54 & 54 & 2  \\
E43.5-M6.5-H3-D8     &  $3\times 10^{43}$ & $3\times 10^{-7}$ & 0.005 &  $10^8$         & 54 & 54 & 3  \\
E43.5-M6.5-H3-D7.7   &  $3\times 10^{43}$ & $3\times 10^{-7}$ & 0.005 &  $5\times 10^7$ & 54 & 54 & 3  \\
E43.5-M6.5-H7-D7     &  $3\times 10^{43}$ & $3\times 10^{-7}$ & 0.005 &  $10^7$         & 54 & 54 & 7  \\
E43.5-M6-H3-D8       &  $3\times 10^{43}$ & $10^{-6}$         & 0.005 &  $10^8$         & 54 & 54 & 3  \\
E44-M6.5-H3-D7.7     &  $10^{44}$         & $3\times 10^{-7}$ & 0.005 &  $5\times 10^7$ & 54 & 54 & 3  \\
E44-M6.5-H3-D8       &  $10^{44}$         & $3\times 10^{-7}$ & 0.005 &  $10^8$         & 54 & 54 & 3  \\
E44-M6.5-H7-D7       &  $10^{44}$         & $3\times 10^{-7}$ & 0.005 &  $10^7$         & 54 & 54 & 7  \\
\hline
\end{tabular}
\end{center}
\end{table*}%

Given the four-fold symmetry of the system, the hydrodynamic equations
are solved in one quadrant of the whole spatial domain in order
to reduce the computational cost. The coordinate system is oriented
in such a way that both the white dwarf and the companion star lie
on the $x$-axis (see Fig.~\ref{fig1}). The companion is at the
origin of the coordinate system, $(x, y, z) = (0, 0, 0)$, and the
computational domain extends to 120~au in the $x$ direction and
60~au in both the $y$ and $z$ directions; the white dwarf is located
to the right on the $x$-axis ($y = z = 0$) at $x = 1.7$~au (namely
the assumed binary separation; see Table~\ref{tab1}). We impose
reflecting boundary conditions at $y\rs{min} = 0$ and $z\rs{min} =
0$ (consistently with the adopted symmetry) and outflow (zero-gradient)
conditions at the other boundaries.

As with previous modeling of other nova blasts (e.g. RS~Oph,
\citealt{2008A&A...484L...9W, 2009A&A...493.1049O}; U~Sco,
\citealt{2010ApJ...720L.195D}; and V407~Cyg, \citealt{2012MNRAS.419.2329O,
2015ApJ...806...27P}), the small scale of the stellar system compared
with the size of the rapidly expanding blast wave over the period
covered presents a major computational challenge. To this end, we
exploit the adaptive mesh capabilities of FLASH by using 10 nested
levels of adaptive mesh refinement, with resolution increasing twice
at each refinement level. The refinement/derefinement criterion
adopted (\citealt{loehner}) follows the changes in mass density,
temperature, and tracer of ejecta. The calculations are performed
using an automatic mesh derefinement scheme in the whole spatial
domain except in the portion including the companion (where we keep
the same resolution of $\approx 0.015$~au during the whole evolution,
corresponding to $\approx 40$ grid points per radius of the companion).
This strategy keeps the computational cost approximately constant
as the blast expands (e.g. \citealt{2015ApJ...810..168O,
2016ApJ...822...22O}): the maximum number of refinement levels used
in the calculation gradually decreased from 10 (initially) to 6 (at
the end) following the expansion of the blast.  At the beginning
(at the end) of the simulations, this grid configuration yielded
an effective resolution of $\approx 0.015$~au ($\approx 0.24$~au)
at the finest level, corresponding to $\approx 20$ zones per initial
radius of the remnant ($> 150$ zones per final radius of the remnant).
The effective mesh size varied from $4096 \times 2048 \times 2048$
initially to $512 \times 256 \times 256$ at the end of the simulation.

From the model results, we synthesize the X-ray emission arising
from the interaction of the blast wave with the surrounding medium.
To this end, we adopt the Astrophysical Plasma Emission Code
(APEC)\footnote{http://www.atomdb.org/} for optically-thin,
collision-dominated plasma with solar abundances of \cite{Anders1989GeCoA}
(AG). We assume abundances of 0.5 times the solar values for the wind
and EDE (\citealt{2015MNRAS.448L..35O}), and AG abundances enhanced
by a factor of 10 for the ejecta. The latter choice was guided by
the evidence of metal-rich ejecta - with abundances enhanced by
possibly more than a factor of 10 - in high-resolution X-ray
spectroscopic studies of RS~Oph and V407~Cyg (e.g.
\citealt{2009ApJ...691..418D, 2011A&A...527A..98S}). The adopted
abundances are important for the estimate of local absorption by
the CSM and ejecta encountered within the blast wave and for the
synthesis of X-ray emission arising from the blast. We adopt the
methodology described by \cite{2009A&A...493.1049O} to synthesize
the X-ray emission. The latter includes the Doppler shift of lines
due to the component of plasma velocity along the line of sight and
the photoelectric absorption by the interstellar medium (ISM), CSM
and ejecta. The absorption is computed using the absorption cross
sections as a function of wavelength from \cite{1992ApJ...400..699B}.
The local absorption is calculated self-consistently from the
distributions of CSM and ejecta; the interstellar absorption is
calculated assuming a neutral hydrogen column density of $N\rs{H}
= 5 \times 10^{21}$~cm$^{-2}$ in agreement with suggestions from
the analysis of {\it Chandra} and {\it Swift} observations
(\citealt{2015MNRAS.454.3108P, 2016ApJ...825...95D}). A distance of $7.8$~kpc
is adopted in agreement with \cite{2010yCat..21870275S}. The exposure
time relevant for capturing the appropriate ``blurring" of the
explosion over the finite observation duration is assumed to be
$t\rs{exp} = 40$~ks (roughly the same order of magnitude of
observations by the {\it Chandra} High Energy Transmission Grating
Spectrometer, HETG; \citealt{2016ApJ...825...95D}).

\section{Results}
\label{s:discuss}

\subsection{Hydrodynamic evolution}
\label{hydro_evol}

\begin{figure*}
  \centering
  \includegraphics[width=17truecm]{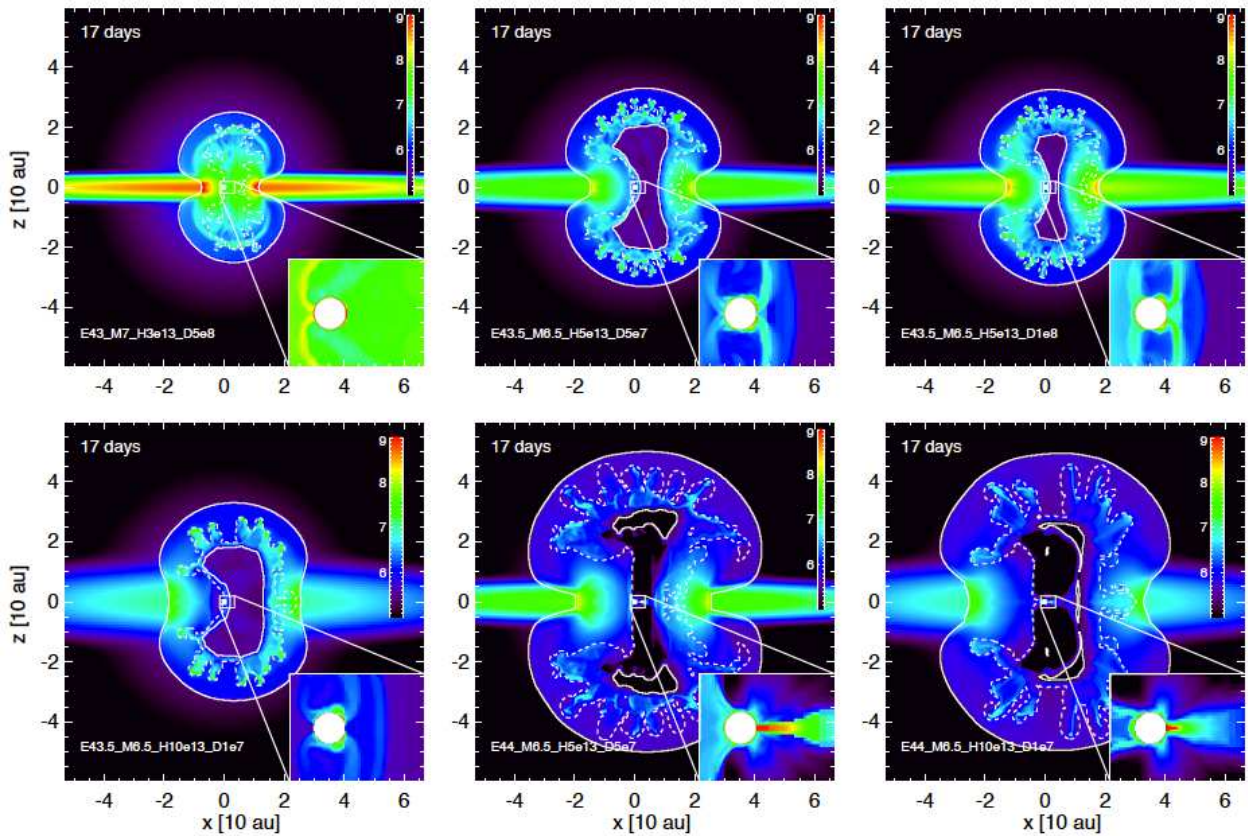}
  \caption{Cross-section images of the gas density distribution,
  on a logarithmic scale in units of cm$^{-3}$ (see the colour table in
  the upper right-hand corner of each panel), at $t = 17$~days since
  the outburst, for six representative V745~Sco models. The red
  giant secondary is at the origin and the initial blast wave is
  offset from the origin to the right by 1.7~au.  Inset panels show
  the blast structure in the immediate vicinity of the binary system.
  The white circle in the inset represents the red giant companion.
  The white dashed contour encloses the ejecta. The white solid
  contours denote the regions with plasma temperature $T > 10^5$~K:
  the outer solid contour marks the position of the forward
  shock, the inner one the position of the reverse shock.}
\label{fig2} \end{figure*}

Our model predicts an evolution of the blast wave which is analogous
to those found for other nova outbursts (\citealt{2008A&A...484L...9W,
2009A&A...493.1049O, 2010ApJ...720L.195D, 2012MNRAS.419.2329O,
2015ApJ...806...27P}). In V745~Sco, the white dwarf is located
within the dense wind of its red giant secondary, as in the case
of RS~Oph and similar to the white dwarf of V407~Cyg which sits in
the massive circumstellar gas envelope of its Mira companion. 
In particular, available information suggests that V745~Sco is
probably analogous to the RS~Oph system, having very similar values
for the binary separation (between 1.5 and 1.7 au), the orbital
period (of the order of 500 days), and the secondary star radius
(between 120 and 150 $R_{\odot}$) (e.g. \citealt{1994AJ....108.2259D,
2000AJ....119.1375F, 2009ApJ...697..721S, 2010yCat..21870275S}).
In both systems the estimated mass accretion rate is of the order
of $\approx 10^{-7}\, M_{\odot}$~yr$^{-1}$ (e.g.
\citealt{2016MNRAS.456L..49K}). On the other hand, V745~Sco has a
much tighter orbital separation than V407~Cyg, implying that, in the
case under study, the nova explodes initially into a relatively
higher density environment. The fact that the white dwarf of V745~Sco
is very close to the companion (with an orbital separation of 1.7~au)
together with the significant size of the companion ($R_* \approx
126\,R_{\odot} \approx 0.6$~AU), implies that the red giant is
expected to shield partially the blast, in analogy with the results
found for the other novae.

Fig.~\ref{fig2} illustrates six representative simulations reproducing
the blast wave evolution of V745~Sco. In particular, the figure
shows the gas density distributions in the $(x, z)$ plane bisecting
the system (the equatorial plane is observed edge-on) at $t =
17$~days, roughly the time when the blast wave was observed by the
{\it Chandra} X-ray observatory. The distribution of ejecta is
delineated by the dashed contour which encloses regions where more
than 90 per cent of the mass is material ejected in the explosion.
As an example, a movie showing the 3D rendering of ejecta density
(in units of cm$^{-3}$) and plasma temperature (in units of K)
distributions during the blast evolution for model E43.5-M6.5-H3-D8
is provided as on-line material\footnote{The blue volume in the
movie is the unshocked EDE with density larger than $10^7$~cm$^{-3}$,
the orange sphere represents the red giant companion, and the white
sphere offset to the right by 1.7~au the initial blast. Note
that the original spatial resolution of the numerical data has been
reduced to save memory when producing the movie describing the full
evolution of the blast in the whole domain.} (movie 1). In
all the cases examined, the system evolution is characterized by
the fast expansion of the shock front with temperatures of a few
millions degrees, and the development of Rayleigh-Taylor (RT)
instabilities at the interface (contact discontinuity) between
shocked ejecta and shocked ambient medium (e.g.
\citealt{1999ApJ...511..335K}).  In particular, the RT instabilities
are responsible for the growth of high density fingers extending
towards the remnant outline (see Fig.~\ref{fig2} and movie 1). The
inner (unshocked) ejecta are cool due to their fast adiabatic
expansion. At this stage, thermal conduction rather than radiative
cooling dominates the evolution of the shock heated plasma. As a
result, both hydrodynamic and thermal instabilities that would
otherwise develop during the blast expansion are significantly
suppressed (e.g.
\citealt{2005A&A...444..505O, 2008ApJ...678..274O}).

\begin{figure}
  \centering
  \includegraphics[width=8truecm]{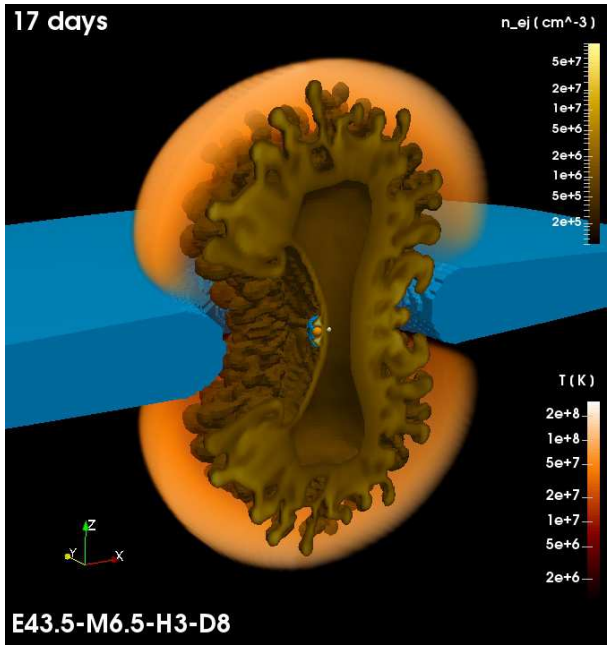}
  \caption{3D rendering of the distributions of ejecta density (in
  units of cm$^{-3}$) and plasma temperature (in units of K), 17~days
  after the outburst for the model E43.5-M6.5-H3-D8. The volume
  has been sliced in the $(x,z)$ plane bisecting the system for
  easy inspection of the blast interior. The blue volume is the
  unshocked EDE with density larger than $10^7$~cm$^{-3}$, the
  orange sphere at the center of the domain represents the red giant
  companion, the white sphere offset to the right by 1.7~au the
  initial blast. The plane of the orbit of the central binary system
  lies on the $(x, y)$ plane. Refer to on-line movie 1 for an
  animation of these data.}
 \label{fig_3d}
\end{figure}

Our simulations show aspherical shock morphologies rendered by the blast
wave propagation through the inhomogeneous circumstellar environment,
similar to those found from the modeling of the RS~Oph and V407~Cyg
outbursts (\citealt{2008A&A...484L...9W, 2009A&A...493.1049O,
2012MNRAS.419.2329O, 2015ApJ...806...27P}). In particular, the
presence of a disc-like EDE leads to a characteristic bipolar shock
morphology in which both the blast and the ejecta are strongly
collimated in polar directions. The collimation is more prominent
for lower explosion energy and/or higher density of EDE (e.g. runs
E43-M7-H2-D8.7 and E43.5-M6.5-H3-D7.7 in Fig.~\ref{fig2}) as might
be expected. Fig.~\ref{fig_3d} shows, as an example, the collimation
of ejecta 17~days after the outburst for the model E43.5-M6.5-H3-D8.
Similar blast collimation but due to the presence of a dense circumstellar
accretion disc was predicted by hydrodynamic simulations of the early
U~Sco blast by \cite{2010ApJ...720L.195D}. The simulations also show
that a reverse shock develops quite early in the evolution as a result of
the interaction of the blast wave with the dense EDE. The reverse shock
travels back into the expanding ejecta, heating them to temperatures
of few MK (lower than the temperature of shocked CSM). We expect,
therefore, that shocked ejecta contribute mainly to X-ray emission at
longer wavelengths.

\begin{figure}
  \centering
  \includegraphics[width=8truecm]{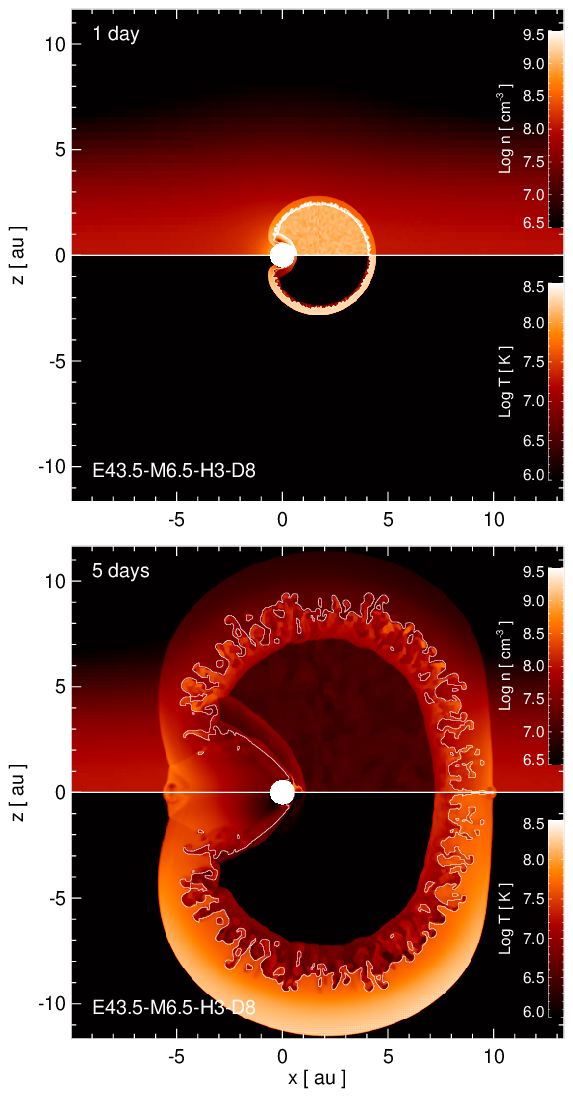}
  \caption{Close-up view of the cross-section images of the gas
  density distribution (upper half-panel) and temperature (lower
  half-panel) on a logarithmic scale (see color tables on the right
  of each panel), at the labeled times for run E43.5-M6.5-H3-D8.
  The white circle at the origin represents the red giant companion;
  the white solid contour encloses the ejecta.}
\label{fig_zoom} \end{figure}

The shape of the blast is also affected by the presence of the red
giant companion. The evolution is similar in all our simulations
and Fig.~\ref{fig_zoom} shows, as an example, the case of run
E43.5-M6.5-H3-D8\footnote{An on-line movie shows a close up view
of the 3D rendering of ejecta density (in units of cm$^{-3}$) and
plasma temperature (in units of K) distributions during the early
evolution of the blast wave for model E43.5-M6.5-H3-D8 (movie 2).
The blue volume is the unshocked EDE with density larger than
$10^8$~cm$^{-3}$ (the plane of the orbit lies on the $(x,y)$ plane)
and the orange sphere represents the red giant companion. Movie
2 shows the evolution of the blast at full spatial resolution.}.
In fact, the shock front propagating towards the companion is
partially shielded and refracted around it. Then the shock follows
an evolution analogous to that found in hydrodynamic simulations
of U~Sco and V405~Cyg: the shock engulfs the companion star and
converges on the rear side of it, undergoing a conical self-reflection
(see lower panel in Fig.~\ref{fig_zoom}). At the same time, a bow
shock with temperatures of few MK and density of $\approx 10^8$~cm$^{-3}$
is produced on the front side of the companion, reheating the ejecta
and contributing to their collimation in polar directions. As
expected the bow shock is more energetic in models with the highest
explosion energy (e.g. runs E44-M6.5-H3-D7.7 and E44-M6.5-H7-D7 in
Fig.~\ref{fig2}). In these cases, the interaction of the shock with
the expanding unshocked ejecta leads to the formation of a dense
region on the front side of the red giant (see lower center and
right panels in Fig.~\ref{fig2}).

\begin{figure}
  \centering
  \includegraphics[width=8truecm]{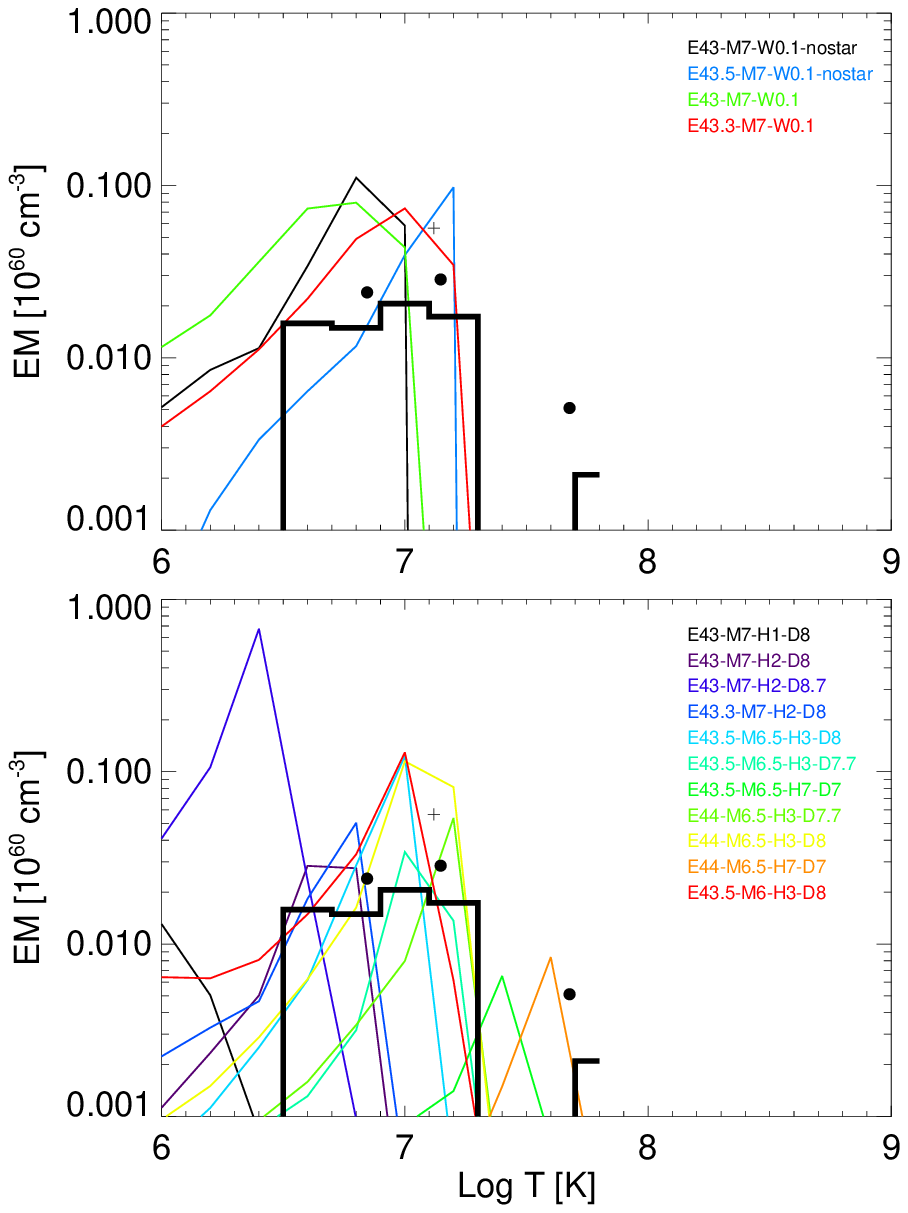}
  \caption{Emission measure vs temperature distributions, EM$(T)$,
  of the blast wave at day 17 derived from models either without
  (upper panel) or with (lower) EDE (see Table~\ref{tab1}). The
  black histogram and the data points represent the EM$(T)$ estimated
  from the analysis of {\it Chandra}/HETG observations with an
  eight-temperature model fit (histogram), with a three-temperature
  model fit (black dots), and with an isothermal model fit (crosses)
  (\citealt{2016ApJ...825...95D}). The factor $10^{60}$ assumes a
  distance for V745~Sco of 7.8~kpc.}
  \label{fig_emt}
\end{figure}

From the simulations, we derive the distribution of emission measure
vs temperature of the blast wave in order to compare our model
results with those obtained from the analysis of {\it Chandra}
observations (\citealt{2016ApJ...825...95D}). From the spatial
distribution of mass density, we first derive the emission measure
in the $j$-th domain cell as em$\rs{j} = n^2\rs{Hj} V\rs{j}$, where
$n^2\rs{Hj}$ is the hydrogen number density in the cell, $V\rs{j}$
is the cell volume, and we assume fully ionized plasma. The EM$(T)$
distribution is then derived by binning the emission measure values
as a function of temperature; the range of temperature $[6 < \log
T ({\rm K}) < 9]$ is divided into 15 bins, all equal on a logarithmic
scale. Figure~\ref{fig_emt} shows the EM$(T)$ for models either
without (upper panel) or with (lower panel) EDE at day 17. The
figure also shows the EM$(T)$ inferred from the analysis of {\it
Chandra}/HETG spectra (\citealt{2016ApJ...825...95D}).

All the models (either with or without the EDE) show a similar trend
of the EM$(T)$ distribution. The shape is characterized by a bump
centered at temperatures between $10^6$ and $10^8$ K, depending on
the parameters of the blast and CSM. The peak of EM is at higher
temperatures for higher values of the explosion energy, $E\rs{0}$,
and/or for lower values of density of the EDE, $n\rs{ede}$. The
peak of EM increases for higher values of EDE density, $n\rs{ede}$,
and/or for higher mass of ejecta, $M\rs{ej}$. Models without EDE
require a high density of the red giant wind, $n\rs{w}$, in order
to fit the EM inferred from the observations at $T\approx 10^7$~K.
Models including an EDE require a less dense wind and predict that
the shape of EM$(T)$ depends on the thickness of the EDE. Among
these, the models best matching the observed EM$(T)$ are those with
explosion energy $E\rs{b}$ between $3\times 10^{43}$ and $10^{44}$~erg,
ejecta mass $M\rs{ej}$ between $3\times 10^{-7}$ and $10^{-6}\,M_{\odot}$,
EDE density $n\rs{ede}$ between $5\times 10^7$ and $10^8$~cm$^{-3}$,
and thickness of the EDE $h\rs{z} \approx 3$~au.
\begin{figure}
  \centering
  \includegraphics[width=8truecm]{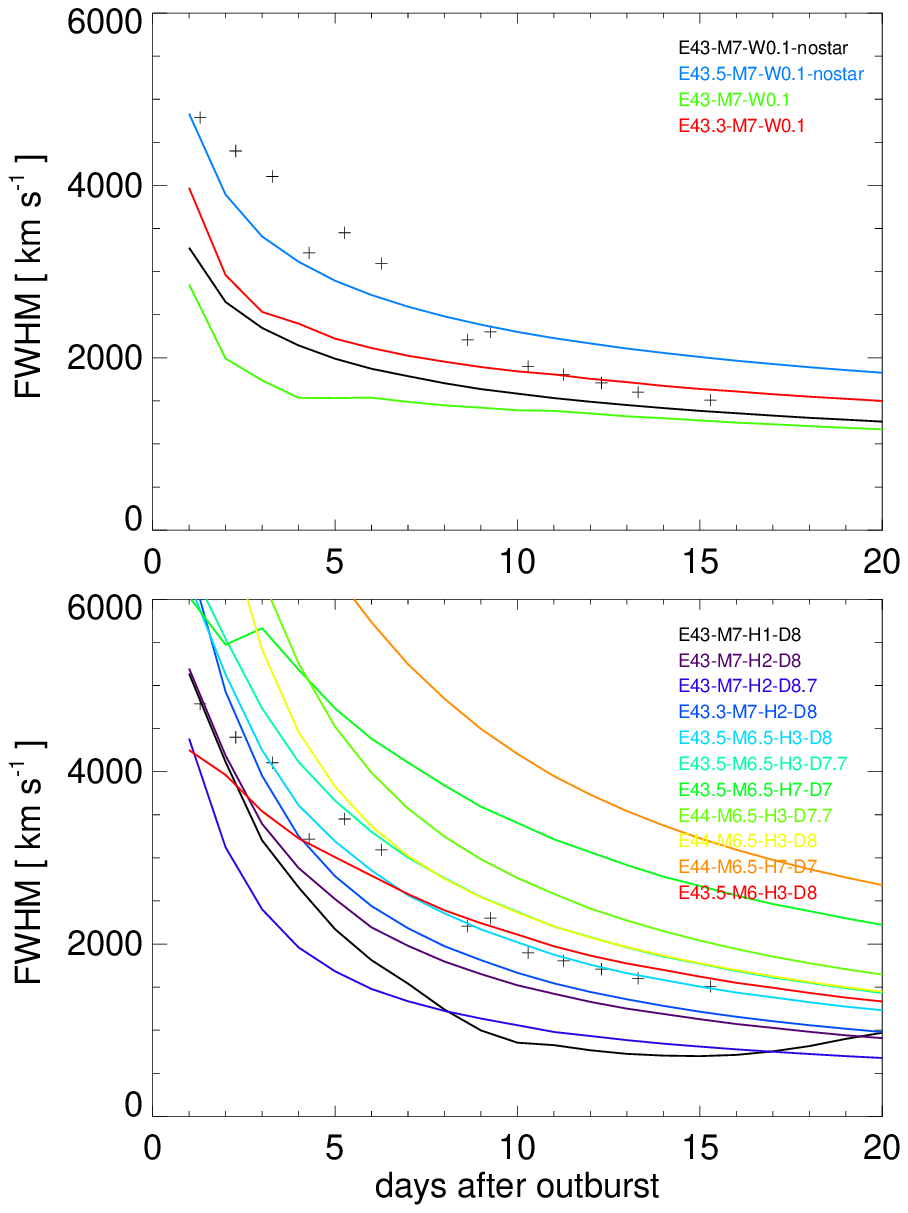}
  \caption{FWHM of H~I Pa$\beta$ emission line vs. time predicted
  by models either without (upper panel) or with (lower) EDE
  (see Table~\ref{tab1}). The data points (crosses) represent the
  values inferred from observations of V745~Sco
  (\citealt{2014ApJ...785L..11B}). The data points correspond to
  the broader line component in the line widths of
  \citet{2014ApJ...785L..11B} which is interpreted as arising
  from the forward shock.}
  \label{fig_vel}
\end{figure}
\begin{figure}
  \centering
  \includegraphics[width=8truecm]{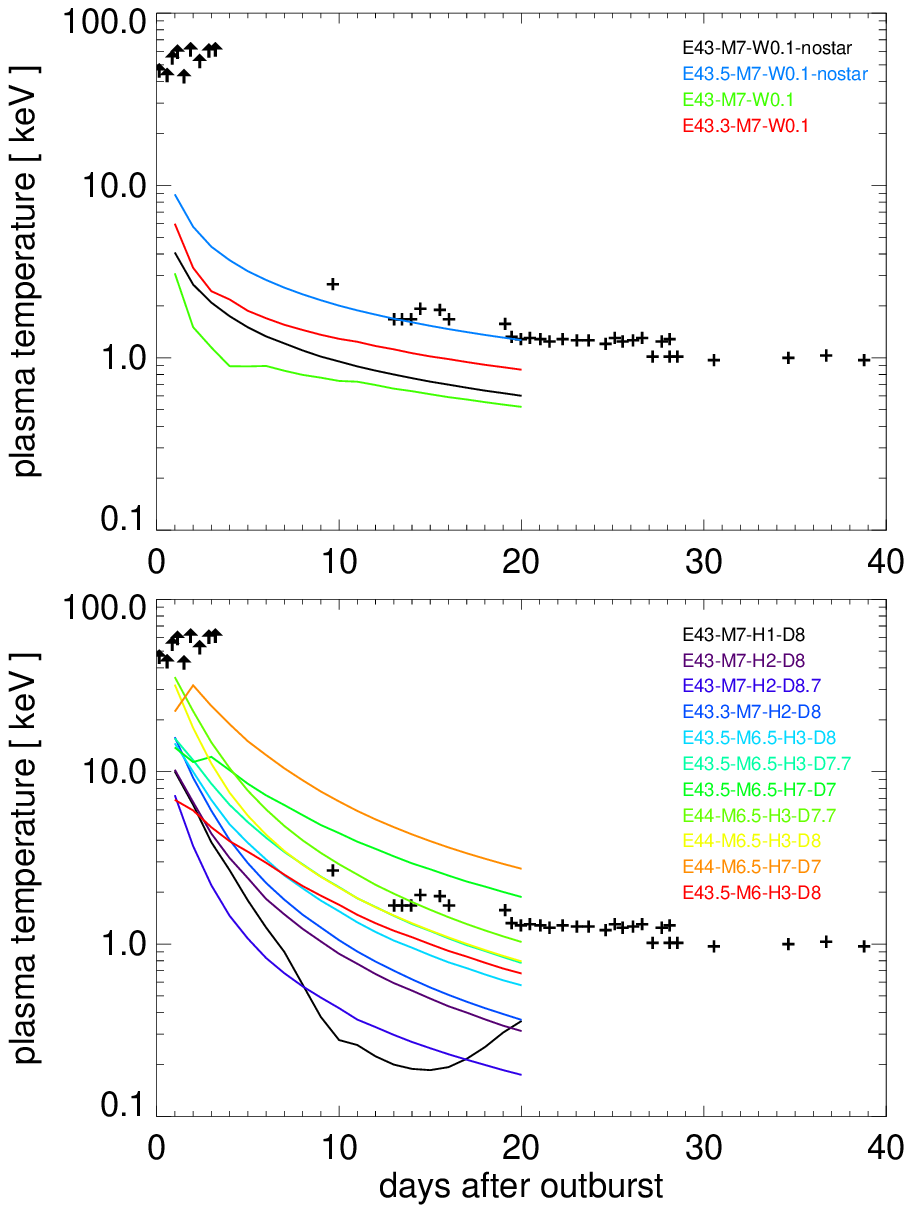}
  \caption{Average plasma temperature vs. time for the V745~Sco
  blast wave predicted by models either without (upper panel)
  or with (lower) EDE (curves; see Table~\ref{tab1}) and those
  inferred from the analysis of {\it Swift} X-ray Telescope
  observations (crosses and arrows; \citealt{2016ApJ...825...95D}).
  The arrows at early times indicate lower limits to the temperature.}
  \label{fig_tem}
\end{figure}

We note that, apparently, the EM$(T)$ inferred from the observations
is characterized by two peaks, one centered at $T\approx 10^7$~K
and the other at $T \approx 5\times 10^7$~K. Our models are able
to reproduce only the low temperature peak. From the analysis of
{\it Chandra} observations of the 2006 outburst of RS~Oph,
\cite{2009ApJ...691..418D} noted that EM values at high temperatures
may be influenced by effects of non-equilibrium ionization (NEI).
However, the highest temperature is mainly determined by the continuum
in the fits of observed spectra of V~745 Sco. The continuum shape
follows the electron temperature and, in fact, \cite{2016ApJ...825...95D}
have found that the fit to {\it Chandra} HETG looks really good out
to the shortest wavelengths (see Fig.~3 of \citealt{2016ApJ...825...95D}).
The Fe~XXV ($\lambda 1.85$) line appears to be slightly underpredicted
by the spectral model, suggesting that Fe ions might suffer from
NEI effects. It is also possible that the hot plasma is a vestige
of the super-hot ($T > 4\times 10^7$~K) plasma observed in the first
three days of the outburst by {\it Swift} (\citealt{2015MNRAS.454.3108P}).
Our models cannot describe this super-hot component and, in fact,
they predict shock temperatures significantly lower than those
observed during the first few days, even with the EDE.

In order to constrain better the model parameters, we compare the
evolution of average velocity and temperature of the blast derived
from the simulations with those inferred from observations. In
particular \cite{2014ApJ...785L..11B} analyzed the profile of H~I
Pa$\beta$ emission line between day 1 and 16 after outburst and
found that, in general, the line is composed of broad and narrow
components. Following \cite{2016ApJ...825...95D}, we interpret the
broad component as arising from the forward shocks, so that its
full width at half maximum (FWHM) should reflect the average velocity
of the blast (\citealt{2014ApJ...785L..11B}; see Fig.~\ref{fig_vel}).
The average temperature of the blast as a function of time was
derived by \cite{2016ApJ...825...95D} from the analysis of {\it
Swift} X-ray Telescope observations (see Fig.~\ref{fig_tem}).  From
the simulations, we derive the velocity, $<v\rs{sh}>$, and temperature,
$<T\rs{sh}>$, of the forward shock, both averaged over the whole
remnant outline and weighted for the emission measure. Then the
FWHM of the H~I Pa$\beta$ line is calculated as $1.8\, <v\rs{sh}>$
(see Sect. 5 in \citealt{2016ApJ...825...95D}).

The comparison between modeled and observed quantities is reported
in Figs.~\ref{fig_vel} and \ref{fig_tem}. Our 3D models without EDE
(upper panels in the figures) produce results analogous to those
of \cite{2016ApJ...825...95D}, based on a 1D analytic model
(\citealt{2003ApJ...597..347L}), if the blast is centered on the
origin of the wind (runs E43-M7-W0.1-nostar and E43.5-M7-W0.1-nostar).
The figures show that the effect of shielding of the blast by the
red giant secondary produces significant changes in the evolution
of $<v\rs{sh}>$ and $<T\rs{sh}>$ (see runs E43-M7-W0.1 and
E43.3-M7-W0.1). None of these models produce a satisfactory description of
the observations.

Among the models including the EDE, run E43.5-M6.5-H3-D8 best matches
the average blast velocity vs. time (but it underestimates the
average blast temperature), whereas run E44-M6.5-H3-D7.7 is the one
best matching the average blast temperature (but it overestimates
the average blast velocity). On the other hand, by comparing the
effective areas of {\it Chandra} and {\it Swift}, we note that the
latter is more sensitive to emission from plasma at high temperature;
in fact, the shock temperature inferred from {\it Swift} observations
around day 17 is slightly higher than that inferred from the analysis
of {\it Chandra} observations (see \citealt{2016ApJ...825...95D}). We argue
therefore that {\it Swift} detects preferentially the hotter plasma
in the blast, namely that which propagates in polar directions. A
compromise between the above two models matching either the shock
velocity or the temperature is given by run E43.5-M6.5-H3-D7.7 which
slightly overestimates $<v\rs{sh}>$ by $\approx 10$\% and underestimates
$<T\rs{sh}>$ by $\approx 50$\%. The model with the highest ejecta
mass ($M\rs{ej} = 10^{-6}\,M_{\odot}$) considered, E43.5-M6-H3-D8,
fits resonably well the average blast velocity (although it
systematically underestimates $<v\rs{sh}>$ during the first 5 days
of evolution), but it underestimates the average blast temperature
as run E43.5-M6.5-H3-D8. All these four models (runs E43.5-M6.5-H3-D8,
E43.5-M6.5-H3-D7.7, E44-M6.5-H3-D7.7, E43.5-M6-H3-D8) fit the
observed EM$(T)$ distribution quite well (see Fig~\ref{fig_emt}).

Finally, we note again that all the models underestimate (even by one order
of magnitude) the early-time {\it Swift} temperatures, although
models with EDE are closer to match than models without EDE (compare
upper and lower panels in Fig.~\ref{fig_tem}). This is mainly due
to the early collimation of the blast by the EDE in polar directions
(see Fig.~\ref{fig_zoom}) which makes the shock propagating poleward
much hotter than the shock front in models without EDE. Our favored
models with explosion energy $E\rs{b} \approx 3\times 10^{43}$~erg
(namely runs E43.5-M6.5-H3-D8 and E43.5-M6.5-H3-D7.7) predict early
temperatures of $\approx 20$~keV, more than a factor of 2 lower
than observed. A better match is obtained by models assuming an
explosion energy $E\rs{b} \approx 10^{44}$~erg although, also in
this case, the temperatures decay very quickly (at odds with
observations) due to the fast adiabatic expansion of the blast.
Presumably models with an even larger explosion energy might reproduce
the observed values. However high explosion energies do not seem
to be realistic in the present case and these models are expected
to fail in reproducing the evolution of shock velocity.

\subsection{X-ray emission and spectral line profiles}

\begin{figure*}
  \centering
  \includegraphics[width=17truecm]{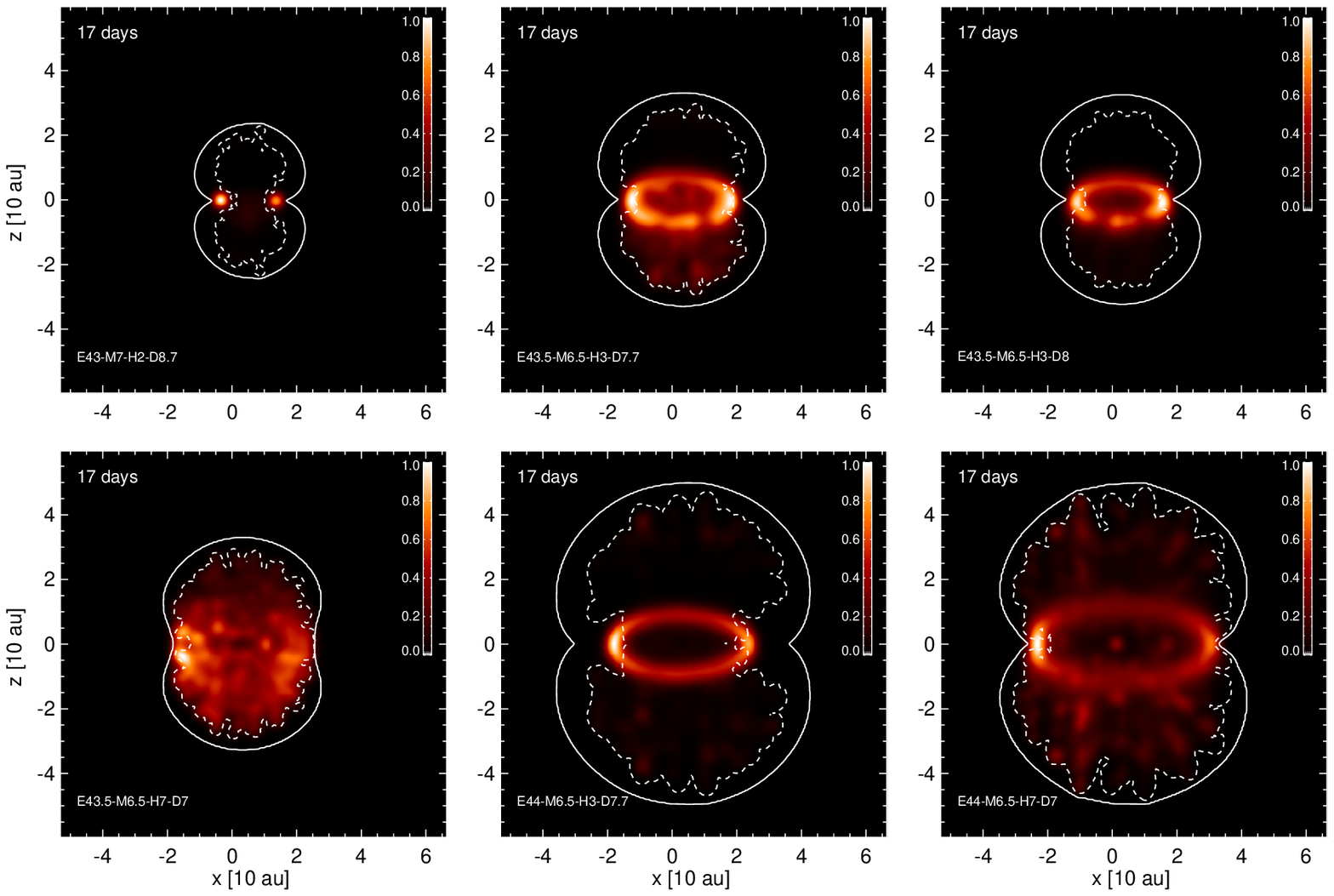}
  \caption{X-ray images of the blast (normalized to the maximum of
  each panel and in linear scale) in the $[0.6 - 12.4]$~keV band
  projected along the line of sight after 17~days of evolution,
  corresponding to the density distributions illustrated in
  Fig.~\ref{fig2}. The plane of the orbit of the central binary
  system lies on the $(x, y)$ plane and is assumed to be inclined
  by $65^o$ relative to the sky plane. Note that, in this figure, $z$
  is the vertical axis in the inclined reference frame. The white
  dashed contour outlines the ejecta projected along the line
  of sight. The white solid contour denotes the projected
  forward shock.} \label{xray_ang25}
\end{figure*}

\begin{figure*}
  \centering
  \includegraphics[width=17truecm]{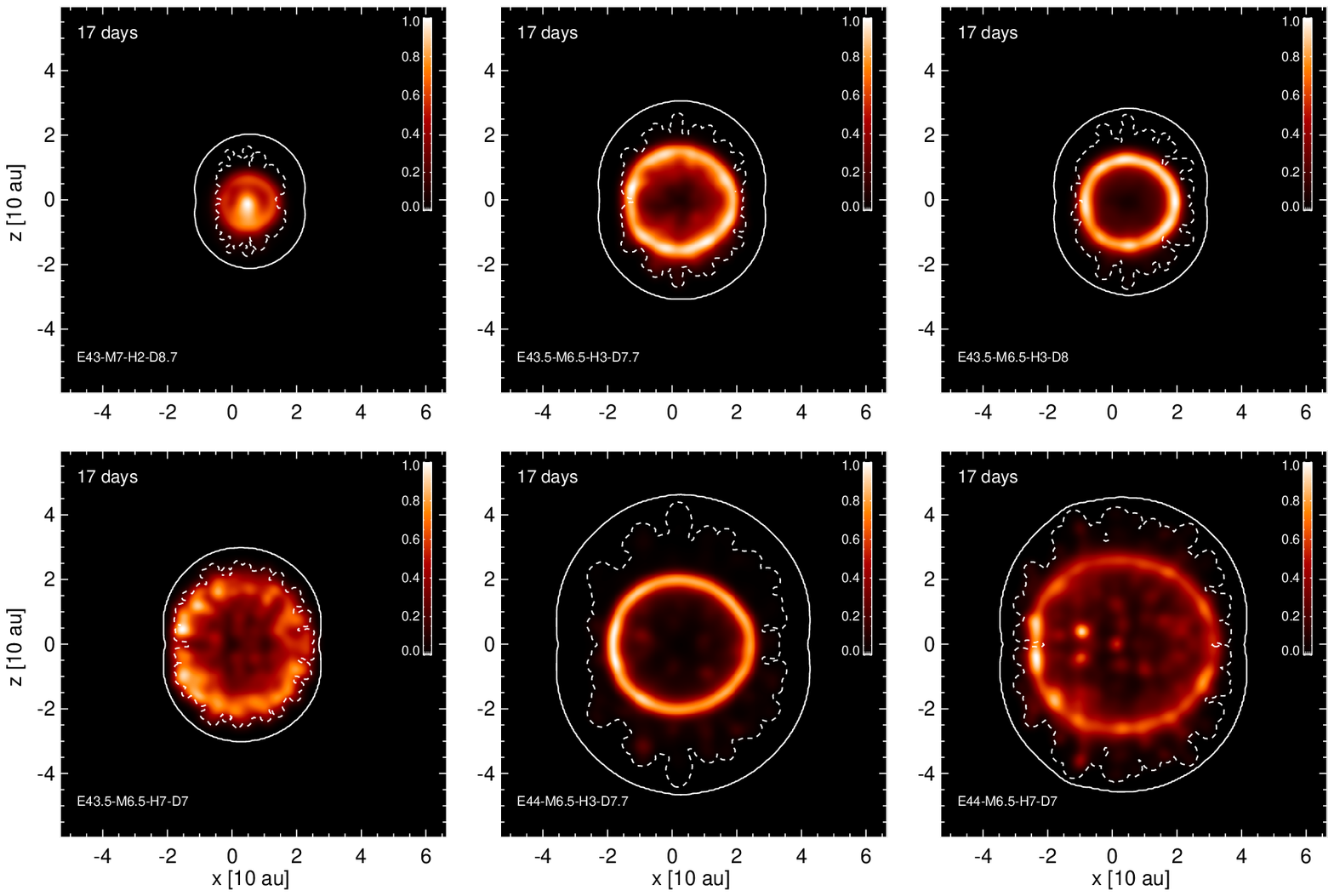}
  \caption{As in Fig.~\ref{xray_ang25} but assuming the plane of the
  orbit of the central binary system to be inclined by $25^o$ relative to
  the sky plane; $z$ is the vertical axis in the inclined
  reference frame.}
\label{xray_ang65} \end{figure*}

From the model results, we synthesize the X-ray emission in the
$[0.6 - 12.4]$~keV band using the method outlined in
Section~\ref{s:obsanal} (see also \citealt{2009A&A...493.1049O} for
more details). Figs.~\ref{xray_ang25} and \ref{xray_ang65} show
maps of X-ray emission projected along the line of sight at day 17
(namely the time of {\it Chandra} observations;
\citealt{2016ApJ...825...95D}) for the models reported in
Fig.~\ref{fig2}. Since the binary inclination remains unknown, in
order to explore the observational consequences of different
inclinations we assume an inclination of the orbital plane relative
to the sky plane either of $65^o$ (Fig.~\ref{xray_ang25}) or of
$25^o$ (Fig.~\ref{xray_ang65}). The white dashed contours in
the figures outline the ejecta projected along the line of sight.
For high inclination of the binary orbit ($i = 65^o$),
we find that, in general, most of the X-ray emission
originates from the interaction of the blast with the EDE and is
concentrated on the equatorial plane. Due to projection effects,
the emission appears either in the form of small-scale
sources propagating in the direction perpendicular to the line-of-sight
(e.g. run E43-M7-H2-D8.7 in the upper left panel of Fig.~\ref{xray_ang25})
or as a ring-like structure lying in the equatorial plane (e.g. run
E43.5-M6.5-H3-D8 in the upper right panel of Fig.~\ref{xray_ang25}).
Similar plasma structures have been found in numerical simulations
describing other nova outbursts (e.g. \citealt{2008A&A...484L...9W,
2009A&A...493.1049O, 2012MNRAS.419.2329O}) and are generally expected
as a result of the propagation of the blast through the EDE. In
other cases, a smaller contribution to X-ray emission arises also
from the ejecta collimated in polar directions (e.g. run E44-M6.5-H7-D7
in the lower right panel of Fig.~\ref{xray_ang25}). In some cases
the contribution from the ejecta is dominant (e.g. run E43.5-M6.5-H7-D7
in the lower left panel of Fig.~\ref{xray_ang25}) and the X-ray
source is a polar cap propagating toward the observer (namely the
portion of the blast less affected by local absorption from dense
ejecta). For low inclination of the binary orbit ($i = 25^o$), the
morphology of the X-ray source is almost ring-like\footnote{
Note that, in run E43-M7-H2-D8.7, the remnant morphology for a high
inclination of the binary orbit ($i = 65^o$; see upper left panel
of Fig.~\ref{xray_ang25}) is characterized by two small-scale sources
instead of a ring-like structure (as for an inclination of $i =
25^o$; upper right panel of Fig.~\ref{xray_ang65}). This is due to
heavy absorption of emission by the dense unshocked EDE for the
proceeding portion of the remnant and by the dense unshocked ejecta
for the receding portion of the remnant.} (see Fig.~\ref{xray_ang65}).
Note that the dashed contours in Fig.~\ref{xray_ang65} outline
the ejecta projected along the line of sight; for low inclination
of the orbit the contours delineate the cross-section of ejecta
located at large distances above (or below) the equatorial plane where
they are more expanded (see Fig.~\ref{xray_ang25}). This may give
the false impression that the emission originates from the ejecta.

\begin{table}
\caption{Parameters (in units of [km s$^{-1}$]) characterizing the
profiles of the most prominent spectral lines in the HEG and MEG
spectra derived from {\it Chandra} observations and from run
E43.5-M6.5-H3-D8, assuming different inclinations $i$ of the orbital
plane relative to the sky plane.}
\label{tab2}
\begin{center}
\begin{tabular}{lrrrrr}
\hline
\multicolumn{6}{c}{model E43.5-M6.5-H3-D8 assuming $i \approx 65^o$} \\
\hline
                    &  $v\rs{ctr}$  & FWHM  &  FWZI  & BSZI   & RSZI \\
\hline
S\,XVI\,($\lambda 4.72$)    &  -25  &  1121 &  2902  & -1476  &  1426 \\
S\,XV\,($\lambda 5.03$)     &   -5  &  1121 &  2882  & -1446  &  1436 \\
Si\,XIV\,($\lambda 6.18$)   &  -35  &  1021 &  2652  & -1366  &  1286 \\
Si\,XIII\,($\lambda 6.64$)  &  -35  &   960 &  2482  & -1276  &  1206 \\
Mg\,XII\,($\lambda 8.42$)   &  -85  &   941 &  2332  & -1256  &  1076 \\
Mg\,XI\,($\lambda 9.16$)    &  -95  &   910 &  2352  & -1266  &  1086 \\
Ne\,X\,($\lambda 12.13$)    & -225  &   940 &  2442  & -1446  &   995 \\
Fe\,XVII\,($\lambda 15.01$) & -295  &  1011 &  2612  & -1606  &  1006 \\
O\,VIII\,($\lambda 18.97$)  & -405  &  1321 &  3423  & -2117  &  1306 \\
\hline
\multicolumn{6}{c}{model E43.5-M6.5-H3-D8 assuming $i \approx 25^o$} \\
\hline
                    &  $v\rs{ctr}$  & FWHM  &  FWZI  & BSZI   & RSZI \\
\hline
S\,XVI\,($\lambda 4.72$)    &  -15  &   880 &  2282  & -1156  &  1126 \\
S\,XV\,($\lambda 5.03$)     &   15  &   850 &  2202  & -1086  &  1116 \\
Si\,XIV\,($\lambda 6.18$)   &  -15  &   800 &  2082  & -1056  &  1026 \\
Si\,XIII\,($\lambda 6.64$)  &   -5  &   700 &  1802  &  -905  &   895 \\
Mg\,XII\,($\lambda 8.42$)   &  -35  &   670 &  1761  &  -915  &   845 \\
Mg\,XI\,($\lambda 9.16$)    &  -35  &   650 &  1681  &  -875  &   805 \\
Ne\,X\,($\lambda 12.13$)    & -145  &   910 &  2362  & -1326  &  1036 \\
Fe\,XVII\,($\lambda 15.01$) & -575  &  1031 &  2682  & -1916  &   765 \\
O\,VIII\,($\lambda 18.97$)  & -1156 &  1981 &  5115  & -3708  &  1506 \\
\hline
\end{tabular}
\end{center}
\end{table}

\begin{figure*}
  \centering
  \includegraphics[width=14.5truecm]{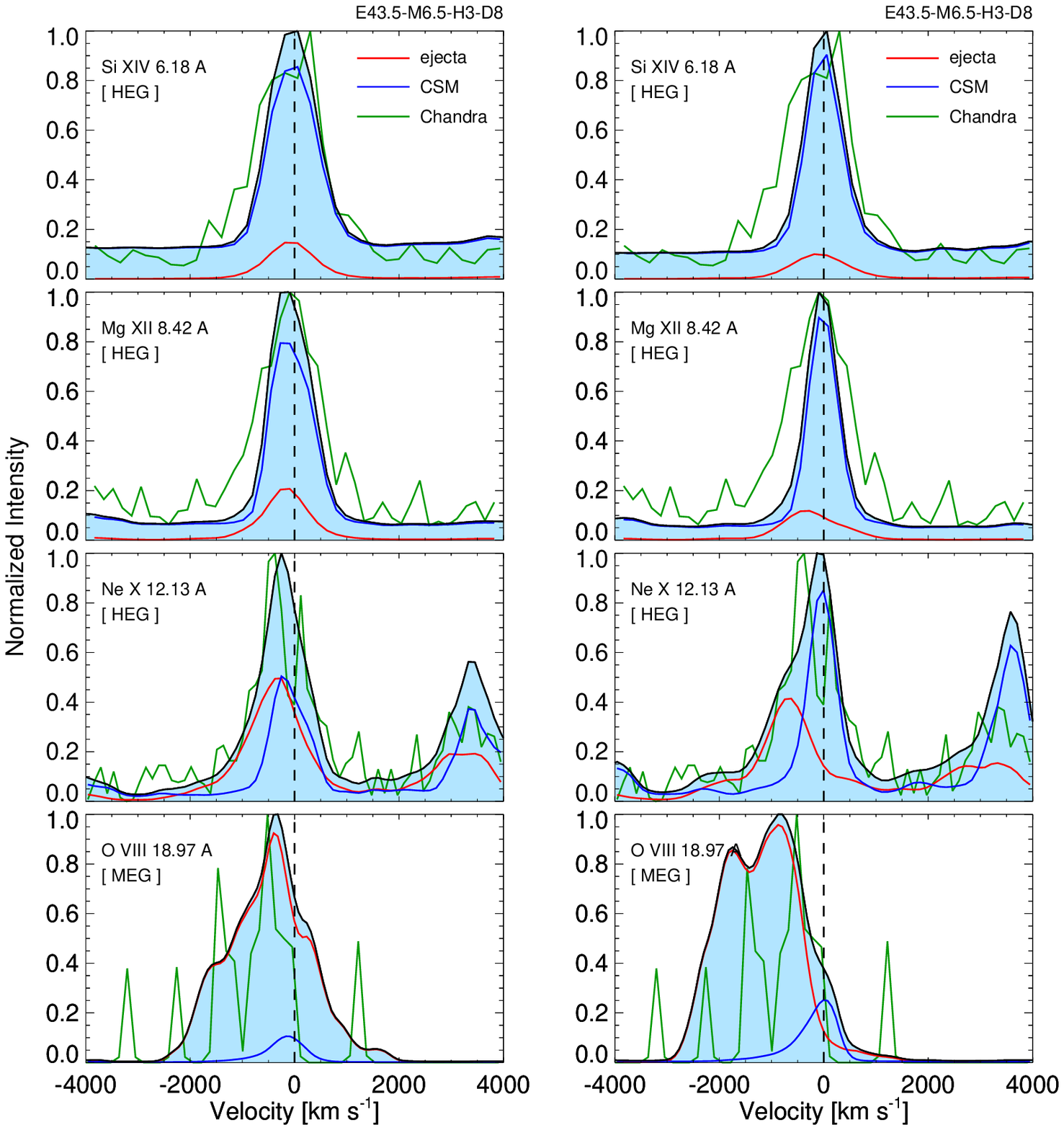}
  \caption{Synthetic velocity profiles of the H-like resonance lines
  of Si\,XIV, Mg\,XII, Ne\,X, and O\,VIII (shades lightblue) derived
  from model E43.5-M6.5-H3-D8 at day 17. The plane of the orbit of
  the central binary system lies on the $(x, y)$ plane and is assumed
  to be inclined either by $65^o$ (on the left) or by $25^o$ (on
  the right) relative to the sky plane. The red and blue lines represent
  the contribution to X-ray emission of shocked ejecta material and
  shocked CSM, respectively. Line profiles observed with the {\it
  Chandra} HETG are superimposed (green lines;
  \citealt{2016ApJ...825...95D}).}
\label{line_prof} \end{figure*}

The synthetic spectra of V745~Sco as predicted to be observed by
the {\it Chandra} High Energy Grating (HEG) and Medium Energy Grating
(MEG) are derived by integrating the emission of the whole spatial
domain. Synthetic line profiles include instrumental broadening
which becomes a more significant component of the line profiles
toward shorter wavelengths.  The synthetic spectra are characterized
by prominent emission lines from different elements, covering a large
range in plasma temperature. This was expected on the basis of the
EM$(T)$ distributions derived from the models, revealing the broad
nature of the plasma temperature distribution (see Fig.~\ref{fig_emt}).
We analyze the synthetic spectra with the aim to investigate the
origin of the broadening and asymmetries detected in the observations
by the {\it Chandra}/HETG (\citealt{2016ApJ...825...95D}). In particular, we
restrict our analysis to the line profiles of the most prominent
spectral lines reported in Table~\ref{tab2}. We find that the model
best reproducing the line profiles of V745~Sco observed with {\it
Chandra} at days 16 and 17 is run E43.5-M6.5-H3-D8. For this model,
Fig.~\ref{line_prof} shows the profiles for the lines selected by
\cite{2016ApJ...825...95D}: the abundant H-like ions Si\,XIV ($\lambda$6.18),
Mg\,XII ($\lambda$8.42), and Ne\,X ($\lambda$12.13) observed by HEG
and O\,VIII ($\lambda$18.97) observed by the MEG\footnote{The O\,VIII
($\lambda$18.97) doublet falls outside of the HEG range and is
observed only by the MEG.}. The figure also compares the synthetic
profiles with those observed with {\it Chandra}
(\citealt{2016ApJ...825...95D}). Note that the signal-to-noise
ratio is rather poor for the O\,VIII line. In the Appendix we
report the line profiles derived for all the models shown in
Figs.~\ref{xray_ang25} and \ref{xray_ang65} and for the model with
the highest ejected mass (E43.5-M6-H3-D8).

The synthetic lines exhibit broadening and asymmetries which are
similar to those observed, especially if the plane of the orbit of
the binary system is inclined by $65^o$ (left panel in
Fig.~\ref{line_prof}). These profiles are also similar to those
predicted by hydrodynamic models describing the 2006 outburst of
RS~Oph (\citealt{2009A&A...493.1049O}). We analyze the line profiles
through fitting with a Gaussian function\footnote{Note that, in
many cases, a Gaussian function is only a rough approximation of
the line profile, because the latter can be asymmetric and consisting
of plasma components with different Doppler shift (see
Fig.~\ref{line_prof}).  Here the fit with a Gaussian component is
intended just to provide indicative values of the line centroid and
broadening.}. Table~\ref{tab2} reports the parameters characterizing
the line profiles for our best-fit case: the shift of the line
centroid $v\rs{ctr}$ (negative values are for blue-shift), the full
width at half maximum (FWHM), the full width at zero intensity
(FWZI), the line profile blue-shift at zero intensity (BSZI), and
the line profile red-shift at zero intensity (RSZI). The values
of the line widths include the instrumental broadening to make
straightforward the comparison of model results with observations.

In velocity terms, the synthetic lines for the case with $i = 65^o$
are characterized by FWZI ranging between $\approx 2300$ and
2900~km~s$^{-1}$ (except the O\,VIII line with FWZI of approximately
3400~km~s$^{-1}$), FWHM ranging between $\approx 900$ and
1100~km~s$^{-1}$, more peaked than expected for a spherically-symmetric
shock, and in agreement with {\it Chandra} observations. The lines
are even more peaked assuming $i = 25^o$ (see Table~\ref{tab2}).
By comparing synthetic and observed line profiles in Fig.~\ref{line_prof},
we note that our most favoured model with $i = 65^o$ slightly
underestimates the observed line widths for the Si\,XIV and Mg\,XII
(left panels in the figure), whereas the Ne\,X line width is very
well reproduced. The line fitting may be improved by changing
slightly the inclination of the orbital plane relative to the sky
plane. On the other hand, it is likely that the imperfect match
of hotter lines is due to simplifications in the prescription of
the EDE and CSM.

\cite{2016ApJ...825...95D} suggested that the pointed shape of
emission lines indicates a highly collimated blast wave and have
shown that line profiles similar to those observed can be produced
if the emission is restricted to an equatorial belt (ring-like)
with a system inclination $i \approx 25^o$ or to one pole (cap-like)
with $i\approx 85^o$. Our model predicts that most of the X-ray emission
originates from the equatorial plane (see Figs.~\ref{xray_ang25} and
\ref{xray_ang65}). In particular the models best matching the EM$(T)$
distribution and the evolution of shock velocity and temperature inferred
from the observations (runs E43.5-M6.5-H3-D7.7 and E43.5-M6.5-H3-D8)
support the scenario of a prominent X-ray emitting equatorial ring,
originating from the interaction of the blast with the EDE.

The centroids of the synthetic lines are, in general, blueshifted
and the amount of the shift depends on the wavelength (see
Table~\ref{tab2} and Fig.~\ref{line_prof}): assuming $i=65^o$ the
net blueshift increases from $\approx -25$~km~s$^{-1}$ in S\,XVI
($\lambda 4.72$) up to $\approx -400$~km~s$^{-1}$ in O\,VIII ($\lambda
18.97$) which is the most striking case. If we assume an inclination
$i=25^o$, again the blueshift increases with the wavelength but now
the trend appears clear only for wavelengths larger than 9~{\AA}; there
the blueshift increases from $\approx -35$~km~s$^{-1}$ in
Mg\,XI ($\lambda 9.16$) up to the very high value of $\approx
-1100$~km~s$^{-1}$ in O\,VIII ($\lambda 18.97$). The result obtained
for $i=65^o$ is that best reproducing the result found from the
analysis of {\it Chandra}/HETG observations. The line profiles tend
to be more extended to the blue than the red, and the effect is
most striking for the O\,VIII doublet (see Fig.~\ref{line_prof}).
As noted by \cite{2016ApJ...825...95D}, this pattern is a clear
signature of intrinsic absorption within the remnant which affects
mainly the emission from the shock heated plasma on the remnant
hemisphere propagating away from the observer. Our simulations
confirm that the prominent blueshift of synthetic lines is caused
by the ejecta which mostly absorb the emission originating from the
receding portion of the remnant that would otherwise contribute to
the redshifted wing of the lines.

The tracer associated with the ejecta allows us to determine the
contribution of shocked ejecta to the X-ray emission. Fig.~\ref{line_prof}
shows that the emission lines consist of two components: one due to
shocked ejecta (red lines in the figure) and the other to shocked CSM
(blue lines). The former increases for longer wavelengths: the emission of
O\,VIII is almost entirely due to shocked ejecta. In fact, as discussed
in Sect.~\ref{hydro_evol}, the interaction of the blast with the CSM
triggers the development of a reverse shock which heats the ejecta to
temperatures lower than that of shocked CSM (see Fig.~\ref{fig_zoom}).
As a result, the dominant contribution to the X-ray spectrum at
wavelengths shorter than $\approx 10$~{\AA} comes from the forward
shock-heated CSM as expected during the early phase in symbiotic
novae (e.g. \citealt{1985MNRAS.217..205B}). This result is also in
agreement with the findings of \cite{2016ApJ...825...95D} that the abundances
derived from the analysis of {\it Chandra} spectra are consistent
with a solar mixture. It is worth mentioning here that our
results depend on the assumption of AG abundances enhanced by a
factor of 10 for the ejecta. A change to the abundances of ejecta
would alter the balance of ejecta vs CSM contributions to emission
lines. This is especially true for the Ne\,X\ line of our best-fit
model in which the two contributions are comparable. On the other
hand, for O\,VIII the ejecta are still expected to be a significant
contributor even for solar-like abundances.

Due to the larger ejecta abundances, the redshifted emission from
the ejecta is, in general, more absorbed than that of shocked CSM.
As a consequence the ejecta component appears more asymmetric and
blueshifted than the CSM component. Inspecting Fig.~\ref{xray_ang25},
we note that a contribution to emission may arise from shocked
ejecta collimated in polar directions (e.g. runs E43.5-M6.5-H3-D7.7
and E44-M6.5-H7-D7 in the figure). The effect is most striking in
run E43.5-M6.5-H7-D7 where the contribution from ejecta dominates
(see also the Appendix). Due to the local absorption, the contribution
to emission is the largest from the ejecta propagating toward the
observer, so that a net blueshift is expected for this component
(see the Appendix for a more detailed analysis of this case). A
larger ejected mass makes the contribution of shocked ejecta to
X-ray emission larger. This is the case, for instance, of run
E43.5-M6-H3-D8 with $M\rs{ej} = 10^{-6}\,M_{\odot}$: the O VIII and
Ne\,X lines are dominated by shocked ejecta (see the Appendix).

Indeed our simulations suggest that an accurate analysis of X-ray
emission lines might reveal useful information about the chemical
composition of the ejecta in these explosions. To do this, we need
to disentangle the ejecta contribution to emission from the CSM
contribution. This depends on the quality of the collected spectra
of course, but it critically depends also on the combination of
explosion parameters and density distribution of the surrounding
CSM. In the case of ``cold'' lines dominated by ejecta, the
poor signal-to-noise ratio obtained for some of these lines (e.g.
see the O\,VIII line in Fig.~\ref{line_prof}) can make difficult
to derive information about the chemical composition of the ejecta.
Nevertheless, in the case of V745~Sco under study, by comparing the
predicted and observed Ne and O spectral line ratios, we find no
signs of strong Ne enhancement that might betray a NeMgO white
dwarf. Thus our model suggests that the progenitor white dwarf is
a CO type.

In the case of hotter lines the ejecta contribution to the total
line profile can be only a small percentage even with abundances
enhanced by a factor of 10; this is the case of Si and Mg in run
E43.5-M6.5-H3-D8 but also of Ne in which the contribution from
ejecta to the total line profile is only about 50\% (see
Fig.~\ref{line_prof}). On the other hand, we find also cases in
which the X-ray emission may be dominated by shocked ejecta at all
wavelengths (see the Appendix for more details). An example is run
E43.5-M6.5-H7-D7 (see lower left panel in Fig.~\ref{xray_ang25})
in which the low density of the EDE makes the contribution of shocked
CSM to emission smaller than that of shocked ejecta. Possibly, in
these cases, we may hope to be able to detect ejecta enriched in
the underlying white dwarf material from blast wave spectra. This
would be valuable for determining more accurately whether the
underlying white dwarf is a CO or an NeMgO type.

\section{Summary and Conclusions}
\label{s:conclusion}

We investigated the origin of broadening and asymmetries of emission
lines observed with {\it Chandra}/HETG during the 2014 outburst of
V745~Sco. The analysis was performed through a 3D hydrodynamic model
which describes the interaction of the blast wave from the outburst
with the inhomogeneous CSM. The model takes into account simultaneously
the radiative cooling and the thermal conduction, and considers an
EDE surrounding the binary system.

We explored the parameter space of the model and found that, in all
the cases, the blast wave is highly aspherical and its morphology
is largely influenced by the pre-existing inhomogeneous CSM. Both
the blast and the ejecta distribution are efficiently collimated
in polar directions due to the presence of the EDE. In addition,
the shock front propagating toward the red giant companion is
partially shielded by it. As a result, depending on the explosion
enegy and density of the EDE, a wake with dense and hot post-shock
plasma can form on the rear side of the companion star.

We searched for the model best fitting the observations by comparing
the average velocity and temperature of the forward shock derived
from the models with those inferred from observations. We found
that the observations are best reproduced if the mass of ejecta in
the outburst was $M\rs{ej} \approx 3 \times 10^{-7}\,M_{\odot}$ and
the explosion energy was $E\rs{b} \approx 3 \times 10^{43}$~erg.
This model predicts a distribution of emission measure vs. temperature
and an evolution of shock velocity and temperature which are
compatible with those derived from the analysis of observations,
and our ejected mass and explosion energy assessments are broadly
in agreement with the estimates of \cite{2014ApJ...785L..11B} and
\cite{2016ApJ...825...95D}.  Interestingly, an ejected mass of
$\approx 3 \times 10^{-7}\,M_{\odot}$ is considerably lower than
the mass needed to initiate the thermonuclear reaction
(\citealt{2016ApJ...825...95D}). If true, then the conclusion is
that the system will be gaining mass making V745~Sco a Type 1a
supernova (SN1a) progenitor candidate.

Our best-fit model allowed us to constrain also the average
density structure and geometry of the pre-nova environment: the
binary system is surrounded by an EDE with density ranging between
$5\times 10^7$ and $10^8$~cm$^{-3}$ and thickness $\approx 3$~au.
This information is important in view of future studies concerning
the origin of the non-thermal emission detected during the blast
wave evolution in radio and $\gamma$-rays (\citealt{2016MNRAS.456L..49K,
2014ATel.5879....1C, cheung15}). In particular the $\gamma$-rays
likely originate in the interaction of the blast and ejecta with
the immediate circumstellar environment (\citealt{2014Sci...345..554A,
2014Natur.514..339C}).  The case of V745~Sco, where the $\gamma$-ray
detection almost coincided with the nova onset, indicates a rapid
particle acceleration and, therefore, suggests a very dense medium
in the immediate vicinity of the white dwarf. This is consistent
with expectations of the structure of the EDE that we have diagnosed
here.

We synthesized the X-ray emission during the blast wave evolution
and found that, in general, most of the X-ray emission originates
from the interaction of the blast with the EDE and is concentrated
on the equatorial plane. Due to projection effects, the emission
appears either in the form of small-scale sources propagating in
the direction perpendicular to the line-of-sight or as a ring-like
structure. A contribution to emission may arise also from the shocked
ejecta collimated in polar directions. In this case the sources of
X-ray emission are, in general, in the form of polar-caps. In all
the cases, the synthetic line profiles are more peaked than expected
for a spherically symmetric shock, in nice agreement with the
observations.

As found from the analysis of {\it Chandra}/HETG observations
(\citealt{2016ApJ...825...95D}), the synthetic line profiles are
asymmetric and slightly blueshifted, especially at wavelengths
larger than 7 {\AA}. Our analysis shows that these asymmetries are
due to substantial X-ray absorption of red-shifted emission by
ejecta material, confirming the conclusion of \cite{2016ApJ...825...95D}.
The X-ray emission lines consist of two components, one originating
from the shocked CSM and the other from shocked ejecta. The
former is the dominant contribution, at least for wavelengths shorter
than $\approx 10$~{\AA}. The latter component suffers more the
effect of local absorption of redshifted emission and exhibits the
largest asymmetries. Our models indicate that, in general, shocked
ejecta contribute substantially to both Ne\,X and O\,VIII lines
(e.g. see Fig.~\ref{line_prof}). While there are still some
discrepancies between models and observations, comparison of predicted
and observed Ne and O spectral line ratios reveals no signs of
strong Ne enhancement (see also the other cases discussed in the
Appendix) that might betray a NeMgO white dwarf and suggests the
progenitor is instead a CO white dwarf. Finally the model best
matching the observed line profiles requires a high inclination of
the orbital plane relative to the sky plane ($\approx 65^o$).

Our simulations confirm that the presence of an EDE or a disc-like
structure around the white dwarf in these systems is an important
ingredient in shaping the expanding blast wave and ejecta distribution
and, thus, in determining the characteristics of the emitted spectra.
Analogous results obtained from the modeling of other recurrent
nova outbursts (RS~Oph, U~Sco, V407~Cyg) suggest that blast and
ejecta collimation by an EDE is likely an ubiquitous feature in these
systems. The results presented here point out once more that the
analysis of X-ray spectra from nova outbursts together with accurate
hydrodynamic modeling may provide information on the structure and
geometry of the environment around these objects and, ultimately,
useful clues to the late stages of stellar evolution.

\section*{Acknowledgments}
This work was partially funded by the PRIN INAF 2014 grant ``Filling
the gap between supernova explosions and their remnants through
magnetohydrodynamic modeling and high performance computing''. SO
thanks his father, Saverio Orlando, for always supporting him and
pushing him towards his dreams. JJD was supported by NASA contract
NAS8-03060 to the CXC and thanks the director, B.  Wilkes, and the
CXC science team for advice and support. The software used in this
work was in part developed by the DOE-supported ASC/Alliance Center
for Astrophysical Thermonuclear Flashes at the University of Chicago.
The simulations were executed at the
SCAN\footnote{http://www.astropa.unipa.it/progetti\_ricerca/HPC/index.html}
(Sistema di Calcolo per l'Astrofisica Numerica) facility for high
performance computing at INAF -- Osservatorio Astronomico di Palermo
(Italy). Finally we thank the anonymous referee whose comments
enabled us to improve the paper.

\bibliographystyle{mn2e}
\bibliography{biblio}

\appendix

\section{Profiles of X-ray emission lines}

\begin{figure}
  \centering
  \includegraphics[width=8truecm]{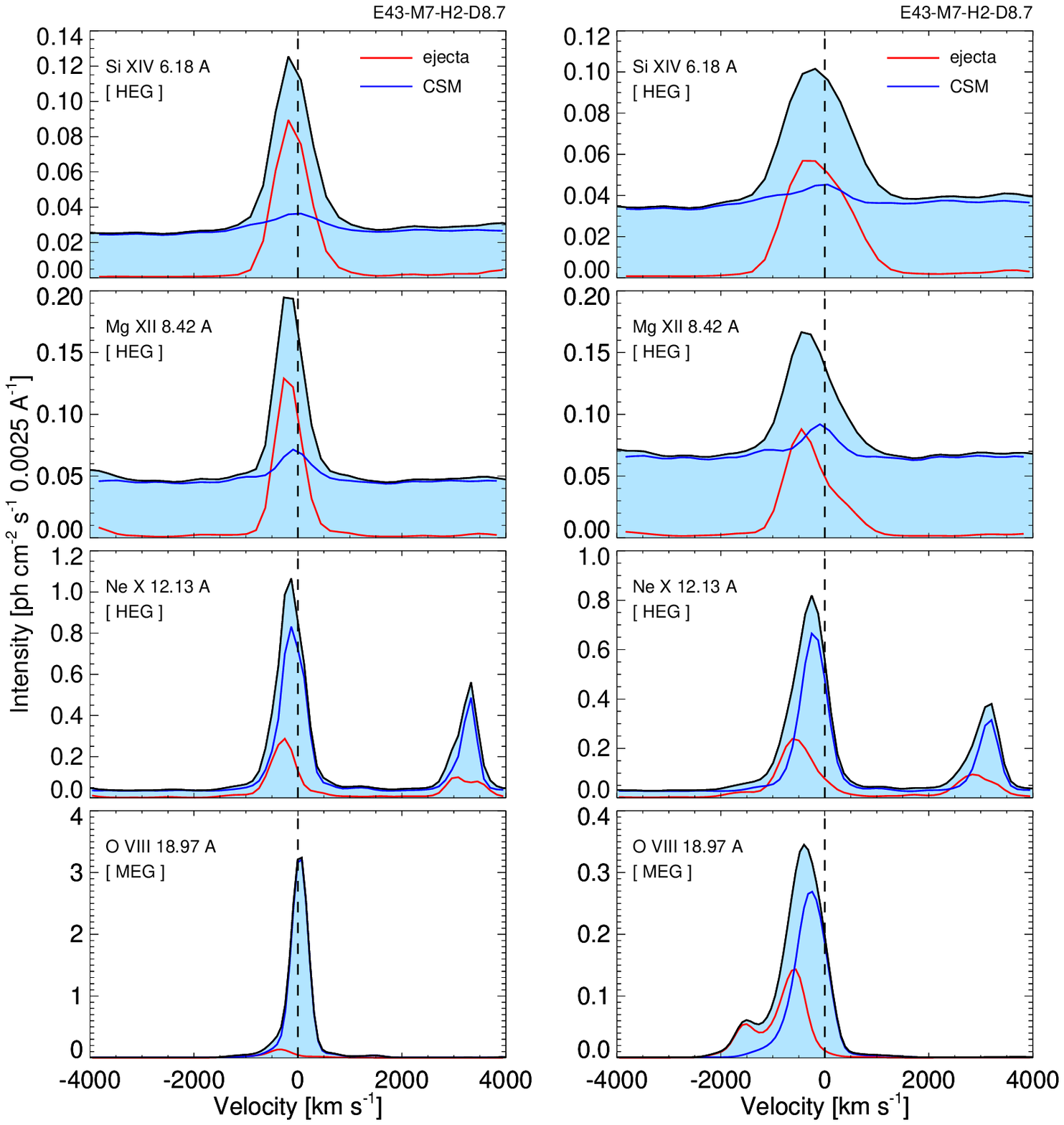}
  \caption{As in Fig.~\ref{line_prof} for run E43-M7-H2-D8.7 (see
  also upper left panel in Figs.~\ref{xray_ang25} and \ref{xray_ang65}).}
\label{E43_M7_H2_D8.7}
\end{figure}

\begin{figure}
  \centering
  \includegraphics[width=8truecm]{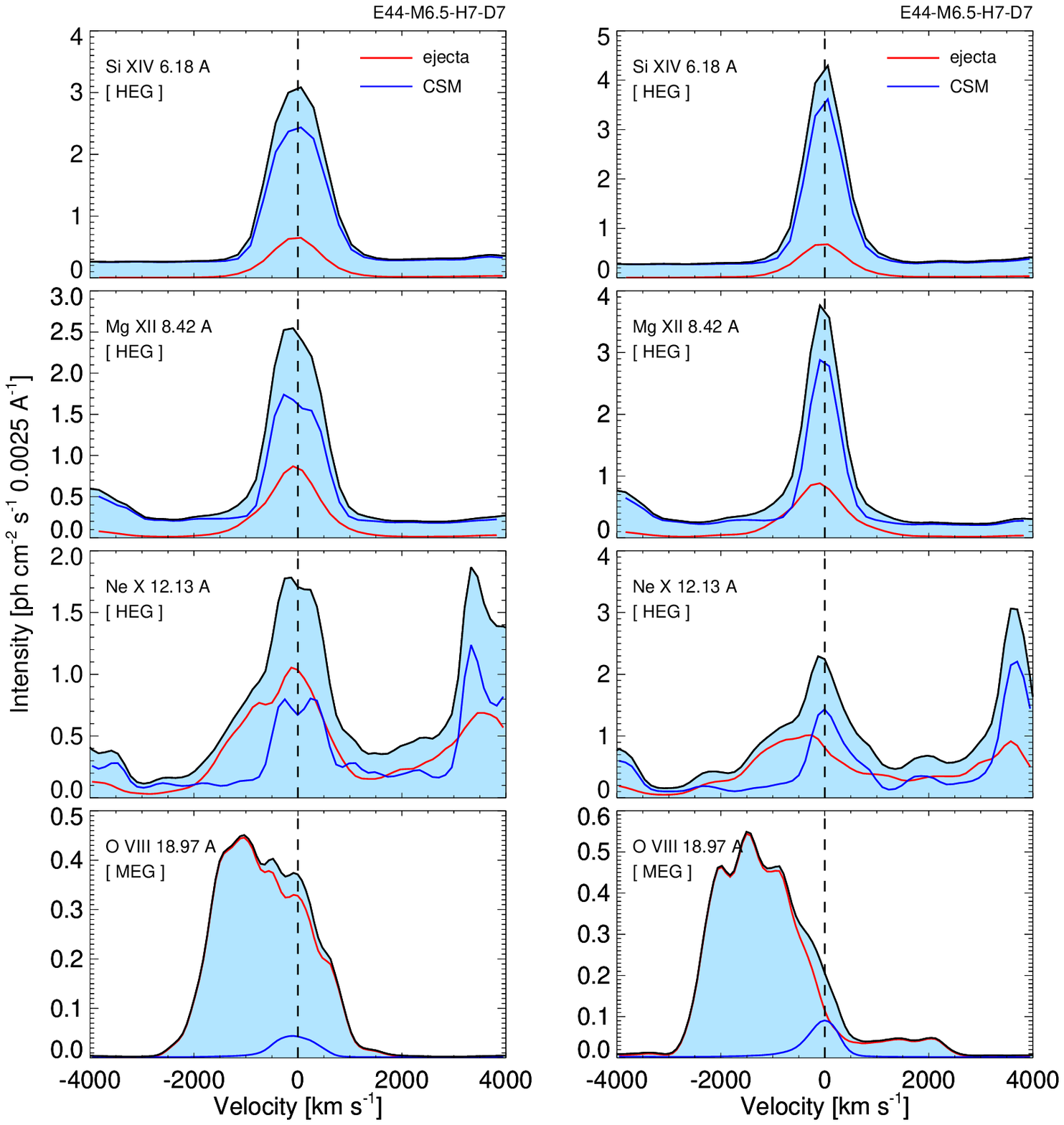}
  \caption{As in Fig.~\ref{line_prof} for run E43.5-M6.5-H3-D7.7 (see
  also upper center panel in Figs.~\ref{xray_ang25} and \ref{xray_ang65}).}
\label{E43.5_M6.5_H3_D7.7}
\end{figure}

\begin{figure}
  \centering
  \includegraphics[width=8truecm]{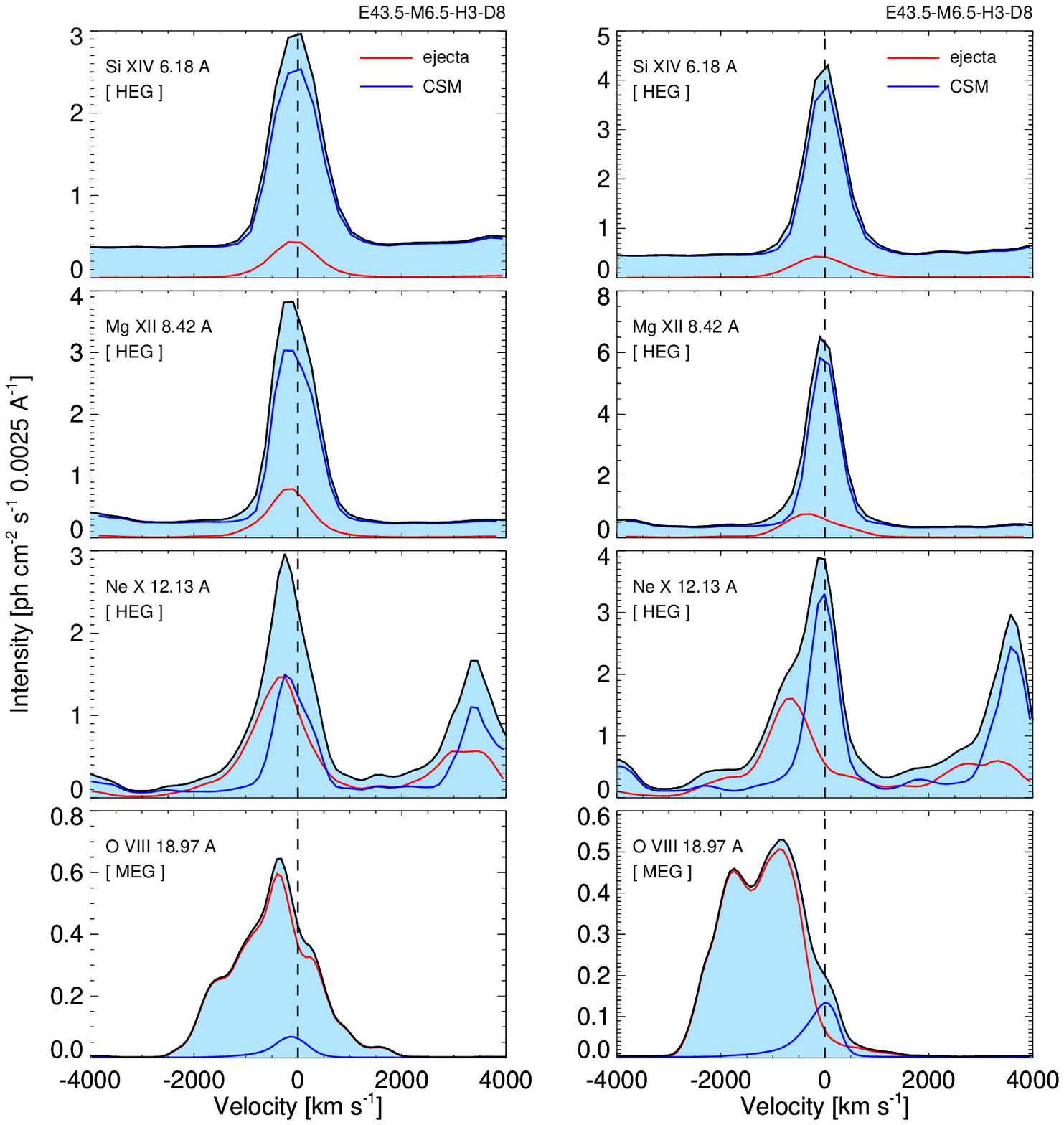}
  \caption{As in Fig.~\ref{line_prof} for run E43.5-M6.5-H3-D8 (see
  also upper right panel in Figs.~\ref{xray_ang25} and \ref{xray_ang65}).}
\label{E43.5_M6.5_H3_D8}
\end{figure}

\begin{figure}
  \centering
  \includegraphics[width=8truecm]{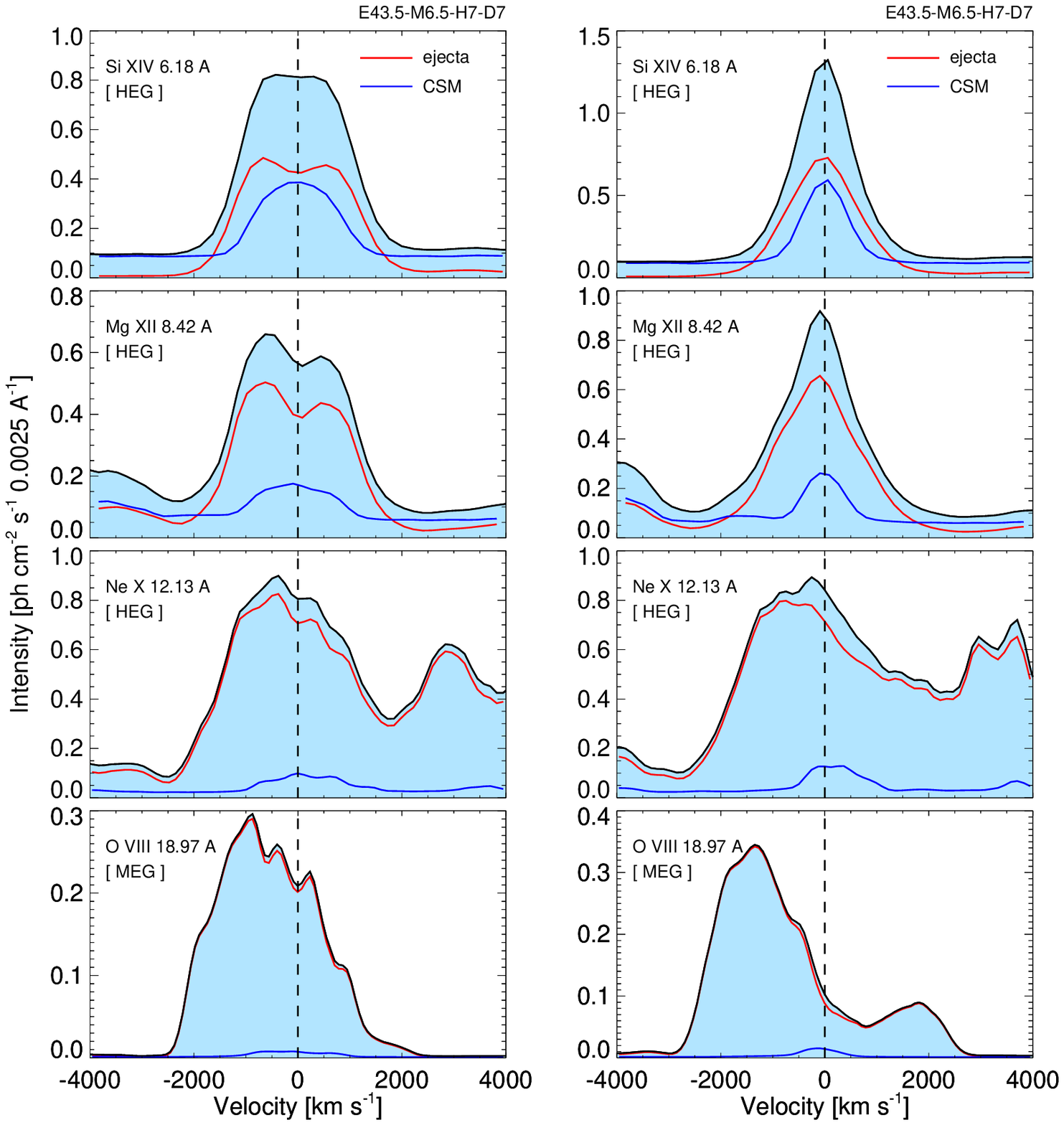}
  \caption{As in Fig.~\ref{line_prof} for run E43.5-M6.5-H7-D7 (see
  also lower left panel in Figs.~\ref{xray_ang25} and \ref{xray_ang65}).}
\label{E43.5_M6.5_H7_D7}
\end{figure}

\begin{figure}
  \centering
  \includegraphics[width=8truecm]{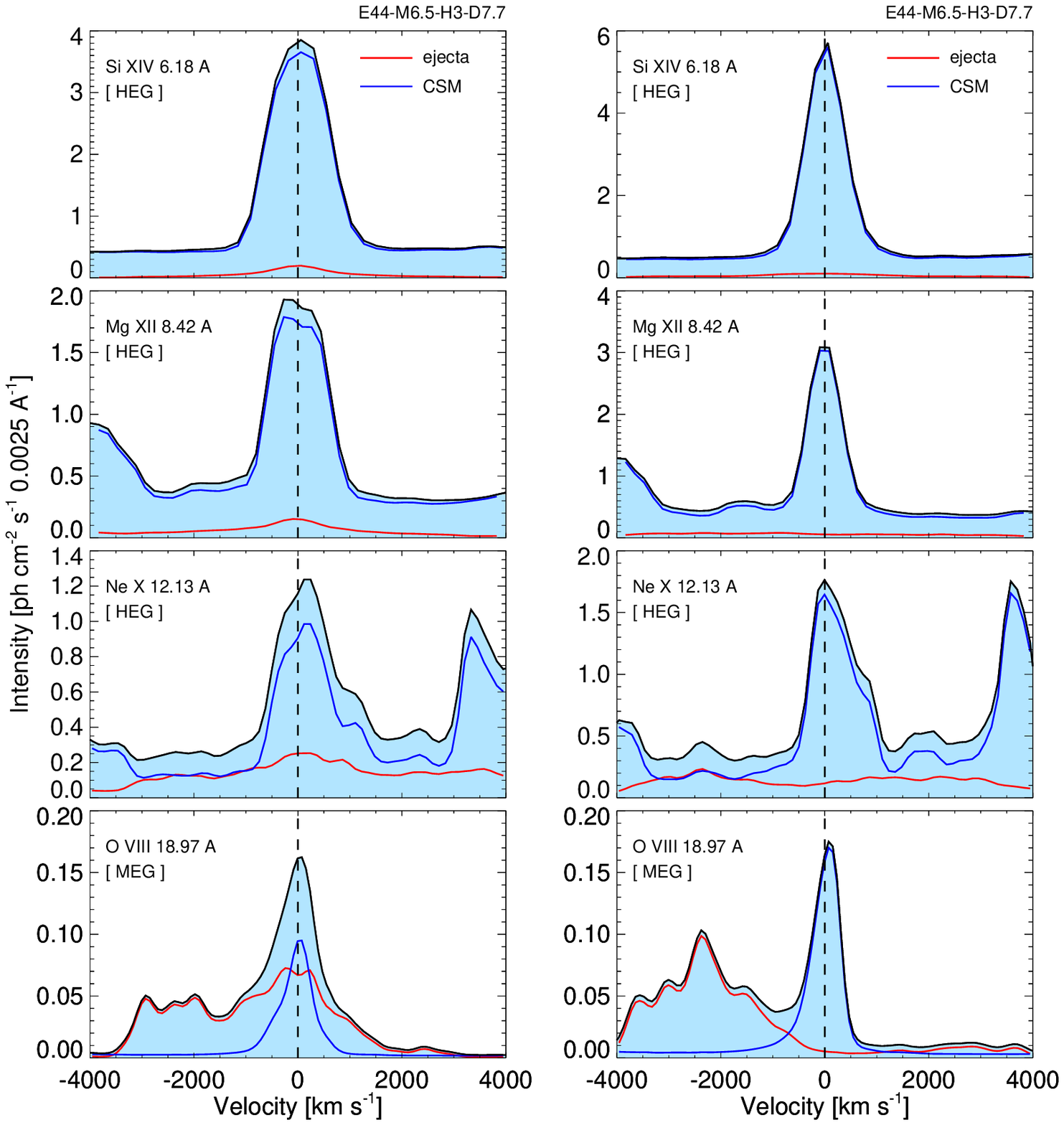}
  \caption{As in Fig.~\ref{line_prof} for run E44-M6.5-H3-D7.7 (see
  also lower center panel in Figs.~\ref{xray_ang25} and \ref{xray_ang65}).}
\label{E44_M6.5_H3_D7.7}
\end{figure}

\begin{figure}
  \centering
  \includegraphics[width=8truecm]{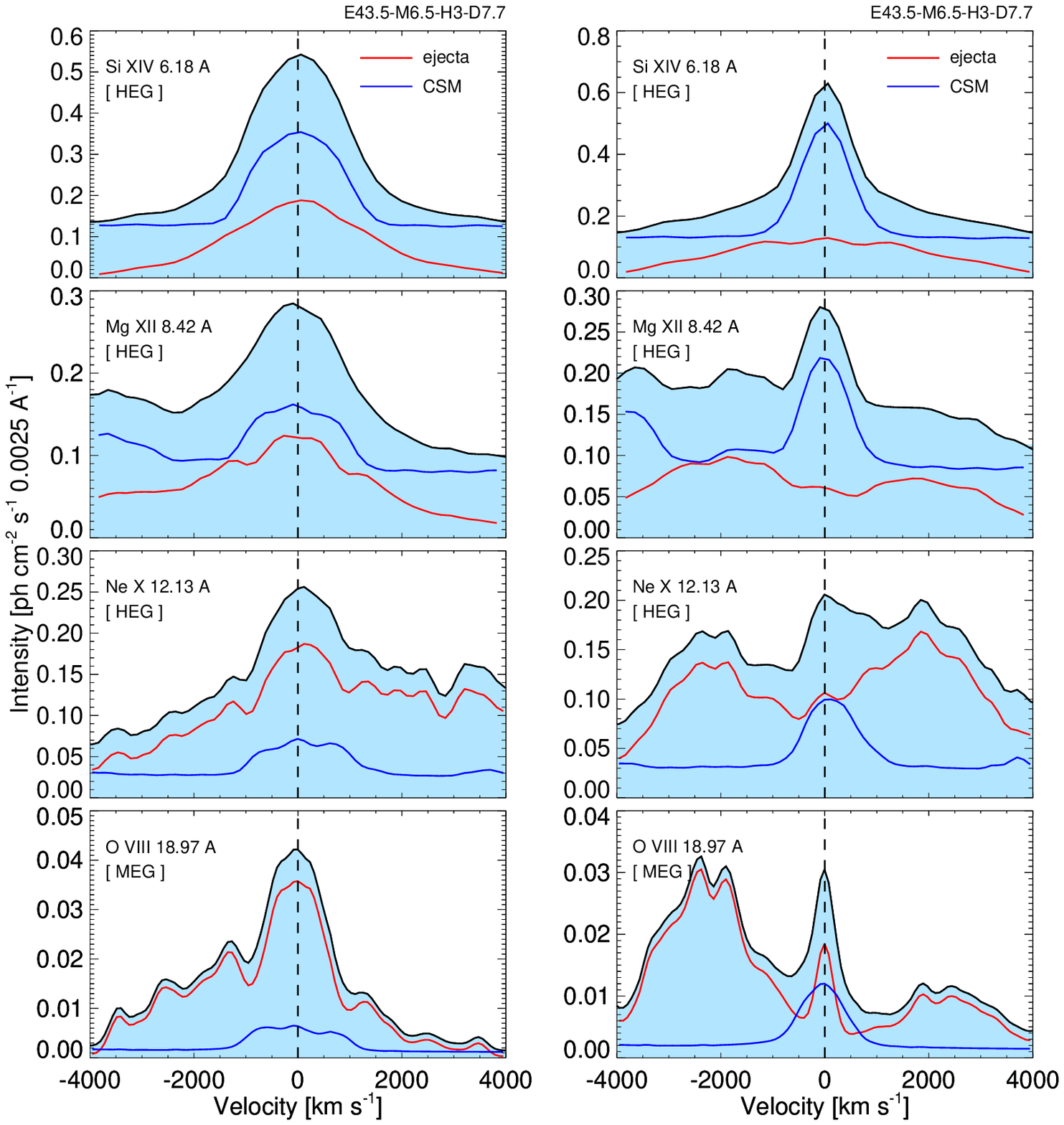}
  \caption{As in Fig.~\ref{line_prof} for run E44-M6.5-H7-D7 (see
  also lower right panel in Figs.~\ref{xray_ang25} and \ref{xray_ang65}).}
\label{E44_M6.5_H7_D7}
\end{figure}

In the paper we discuss in detail the synthetic line profiles as
predicted to be observed by the {\it Chandra} HEG and MEG for our
best-fit model, namely run E43.5-M6.5-H3-D8 (see Fig.~\ref{line_prof}
and Table~\ref{tab2}). Here we report the line profiles of the
abundant H-like ions Si\,XIV ($\lambda$6.18), Mg\,XII ($\lambda$8.42),
Ne\,X ($\lambda$12.13) and O\,VIII ($\lambda$18.97) for the 
models shown in Figs.~\ref{xray_ang25} and \ref{xray_ang65} (see
Figs.\ref{E43_M7_H2_D8.7}-\ref{E44_M6.5_H7_D7}) and for the model
with the highest ejected mass (E43.5-M6-H3-D8). We find that the
models show significant differences in the synthetic line profiles
depending on the combination of explosion parameters and density
distribution of the surrounding CSM.

In general, the emission is dominated by shocked CSM at shorter
wavelengths and by shocked ejecta at longer wavelengths as found
for run E43.5-M6.5-H3-D8. There are however two exceptions. In run
E43-M7-H2-D8.7, the line profiles are very sharp and the emission
is dominated by shocked ejecta at shorter wavelengths and by shocked
CSM at longer wavelengths (see Fig.~\ref{E43_M7_H2_D8.7}). This is
mainly due to the high density of the EDE which makes the reverse
shock heating the ejecta stronger and the shocked CSM denser and
colder than in other models. The result is that the shocked ejecta
are hotter than the shocked CSM so that they contribute more to the
emission of ``hot'' lines.

The other exception is run E43.5-M6.5-H7-D7 in which the main
contribution to line emission originates from shocked ejecta for
all the lines selected (see Fig.~\ref{E43.5_M6.5_H7_D7}). Now the
cause is the low density of the EDE which makes the contribution
to emission from shocked CSM smaller than that in the other models.
We note also that, in this case, the line profiles may show two
peaks (for instance in the Mg\,XII line in the left panel of
Fig.~\ref{E43.5_M6.5_H7_D7}) when the orbital plane is highly
inclined relative to the sky plane ($65^o$). This feature is the
consequence of the collimation of shocked ejecta in polar directions,
so that the sources of X-ray emission are polar-caps propagating
one toward, and the other away from, the observer. As a result, the
double-peaked lines reflect these two ejecta components, one
blueshifted and the other redshifted, with the latter more absorbed
by the dense ejecta and CSM placed along the line-of-sight.

\begin{figure}
  \centering
  \includegraphics[width=8truecm]{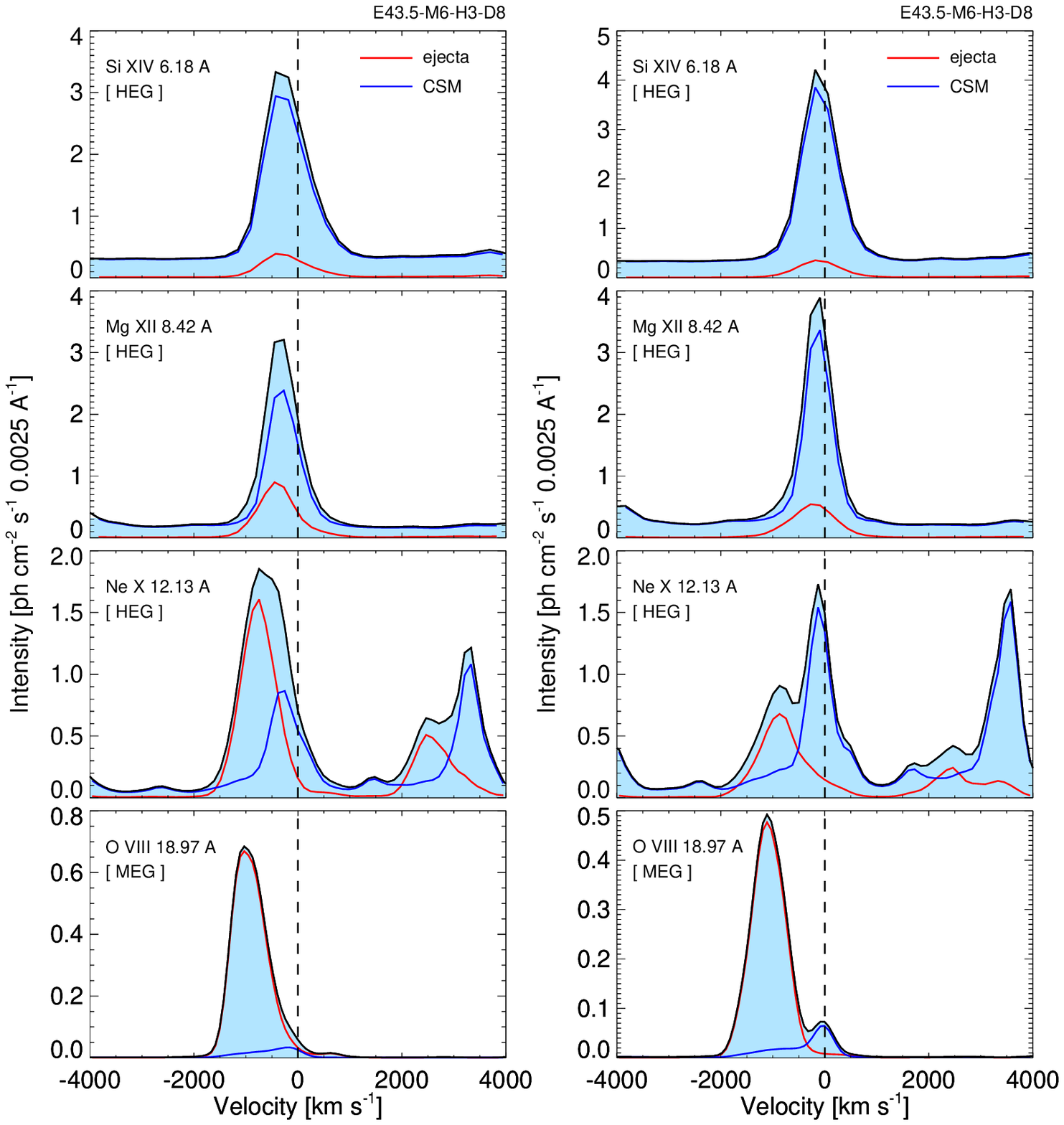}
  \caption{As in Fig.~\ref{line_prof} for run E43.5-M6-H3-D8.}
\label{E43.5_M6_H3_D8}
\end{figure}

Finally, the model with the highest ejected mass (E43.5-M6-H3-D8)
present a significant contribution of shocked ejecta especially at
longer wavelengths (see Fig.~\ref{E43.5_M6_H3_D8}). This is not
surprising because, in this case, the density of ejecta is higher
than in our best-fit model E43.5-M6.5-H3-D8. Since the effect of
local absorption is larger for the ejecta (due to their larger
abundances), the redshifted emission from the ejecta is more absorbed
than that of shocked CSM and the line profiles in run E43.5-M6-H3-D8
are more asymmetric and blueshifted than in run E43.5-M6.5-H3-D8
and in the observations (see \citealt{2016ApJ...825...95D}).

\label{lastpage}

\end{document}